\documentclass[%
 aip,
 amsmath,amssymb,
 reprint,%
]{revtex4-1}

\usepackage{graphicx}
\usepackage{dcolumn}
\usepackage{bm}

\usepackage[utf8]{inputenc}
\usepackage[T1]{fontenc}
\usepackage{mathptmx}
\usepackage{units}
\usepackage{upgreek}
\usepackage{textcomp}
\usepackage[absolute,overlay]{textpos}
\usepackage{helvet}
\usepackage{color}
\begin{document}


\title[Ultrafast Laser Modified C-shaped Glass Edges]{Protecting the Edge: \\ Ultrafast Laser Modified C-shaped Glass Edges}

\author{Daniel Flamm}
\email{daniel.flamm@trumpf.com}
\author{Myriam Kaiser}
\author{Marvin Feil}%
\author{Max Kahmann}%
\author{Michael Lang}%
\author{Jonas Kleiner}%
\author{\mbox{Tim Hesse}}%
\affiliation{$^{1}$TRUMPF Laser- und Systemtechnik GmbH, Johann-Maus-Str. 2, 71254 Ditzingen, Germany
}%




\begin{abstract}
A procedure and optical concept is introduced for ultrashort pulsed laser cleaving of transparent materials with tailored edges in a single pass. The procedure is based on holographically splitting a number of foci along the desired edge geometry including C-shaped edges with local $45^{\circ}$ tangential angles to the surface. Single-pass, full-thickness laser modifications are achieved requiring single-side access to the workpiece only without inclining the optical head. After having induced laser modifications with feed rates of $\sim\unit[1]{m/s}$ actual separation is performed using a selective etching strategy. \\ \\
Submitted: 27 October 2021. Accepted: 5 December 2021. Published online 27 December 2021.
\textsf{This article may be downloaded for personal use only. Any other use requires prior permission of the author and AIP Publishing. This article appeared in J.~Laser Appl.~\textbf{34}, 012014 (2022) and may be found at \url{https://doi.org/10.2351/7.0000592}}.
\end{abstract}

\maketitle

\begin{textblock*}{17cm}(2.25cm,26.75cm) 
   \centering \small 
   \textsf{
   Flamm, D., \textit{et al.}, J.~Laser Appl.~\textbf{34}, 012014 (2022). \url{https://doi.org/10.2351/7.0000592}.}
\end{textblock*}

\section{\label{sec:level1}Introduction}
Strategies for ultrashort pulse laser cutting of sheetlike transparent materials were successfully transferred to industry within recent years. Reliable ultrafast laser platforms are equally available as adapted processing optics providing customized nondiffracting beams for single-pass operations with m/s-feed rates.\cite{jenne2020facilitated}

The substrates with straight face edges, thus fabricated, are susceptible to cracking and chipping in the event of an impact as stress is accumulated at the $90^{\circ}$ corners---a process that the reader may have already observed on his or her own smartphone. Substrates with reduced edge angles, as known from chamfered, beveled or C-shaped edges, will therefore protect the glass substrate by reducing cross sections and by improving impact resistance.\cite{marjanovic2019edge} After a glass substrate has been cut to size, various further process steps may follow, such as coating, hardening, or bonding. Most likely, this will increase the applied mechanical stress. For the mentioned postprocessing steps a chamfered glass substrate is, thus, highly desired. A laser-based machining process will offer added value compared to classical grinding and polishing steps,\cite{bukieda2020study} if cutting and chamfering are achieved within a single operation.

Nondiffracting beams as subtle laser tools for single-pass cutting of transparent materials have been recently used to generate beveled\cite{jenne2018high} or chamfered\cite{ungaro2021using} edges. Here, most intuitively, the processing optics is inclined with respect to the workpiece,\cite{flamm2021ultrafast} requiring sensitive phase corrections to the perturbed optical field.\cite{jenne2018high, cheng2020aberration, ungaro2021using} Besides serious disadvantages in machining, such as the need for a precise five-axis system\cite{flamm2021ultrafast} or a vanishingly small working distance, there are simple optical reasons that prevent the production of shaped edges with $45^{\circ}$ angles using these concepts. Light's critical refraction angle at the tilted glass interface amounts to $\sim$ $40^{\circ}$ (at inclination angle $90^{\circ}$ and refractive index $n=1.5$).\cite{ungaro2021using} As the nondiffracting beam is fed by field components propagating along the cone-angle toward the optical axis,\cite{mcgloin2005bessel} the maximum inclination angle is further reduced, resulting in highest feasible chamfer angles that amount to $\sim$ $30^{\circ}$.\cite{jenne2018high, ungaro2021using}

We will, therefore, introduce a novel concept based on holographic 3D-beam splitters enabling volume deposition of energy along arbitrary edge contours including geometries that yield C-shaped edges.\cite{flamm2021structured} In the present case, \textcolor{black}{using the 3D-beam splitters,} the associated tangential angles of the edge contour to the surface can be chosen arbitrarily and amount to even less than $30^{\circ}$ at both glass surfaces. The processing optics realizing our multispot concept operates under vertical incidence and requires single-side access to the workpiece only (no workpiece flipping required). The ability to introduce full-thickness volume modifications along the entire substrate edge geometry within a single pass allows for laser processing in the order of $\unit[1]{m/s}$. Here, the 3D-beam splitter realized by flexible digital holograms or static diffractive optical elements (DOEs) is designed to exploit the entire power performance of industry-grade ultrashort pulse laser systems.\cite{flamm2021structured} In fact, required laser parameters (pulse energy and duration, repetition and resulting feed rate) are very similar to conventional glass cutting processes based on the usage of nondiffracting beams.\cite{jenne2020facilitated, flamm2021structured}

The laser-based modification step is followed by the actual separation performed chemically in the present study. Here, crack-connected type-III-regime modifications\cite{itoh2006ultrafast} enable selective etching along the desired edge geometry with high etch rates.\cite{kaiser2019selective, rave2021glass} Generated tailored edges meet the demand of the consumer electronics industry regarding edge roughness, stability and throughput.

Although we only demonstrate separation by etching here, other separation strategies are conceivable, too, in particular thermal separation by applying CO$_2$ laser radiation.\cite{nisar2013laser} 

\section{\label{sec:Depo}Energy deposition along arbitrary edge contours}
\begin{figure*}
    \centering
    \includegraphics[width=1\textwidth]{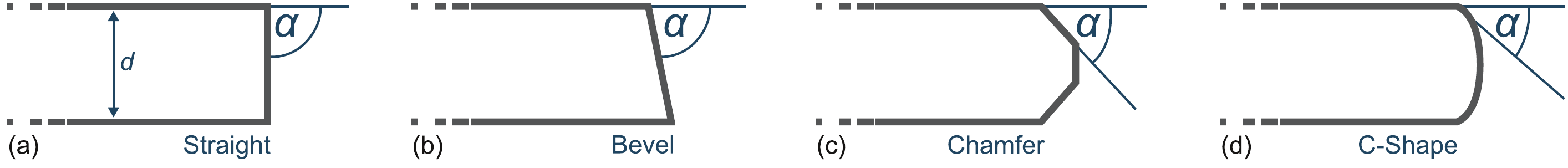}
    \caption{Schematic of different edge geometries applied to transparent sheetlike substrates of thickness $d$ with definition of global or local (tangential) edge angles $\alpha$. (a) Straight, (b) beveled, (c) chamfered, and (d) C-shaped edge.}
    \label{fig:champion}
\end{figure*}

The different types of bulk modifications in transparent materials induced by ultrashort laser pulses\cite{itoh2006ultrafast} and their beneficial use for efficient processing strategies like cutting or welding are well known and understood. The challenge in the present case, i.e. to deposit energy along arbitrary volume contours, is more of an optical question. Is there an optical concept in which a single focus can take the desired form of an edge with reduced edge angle $\alpha$, cf.~Fig.\,\ref{fig:champion}?

\begin{figure}
    \centering
    \includegraphics[width=0.48\textwidth]{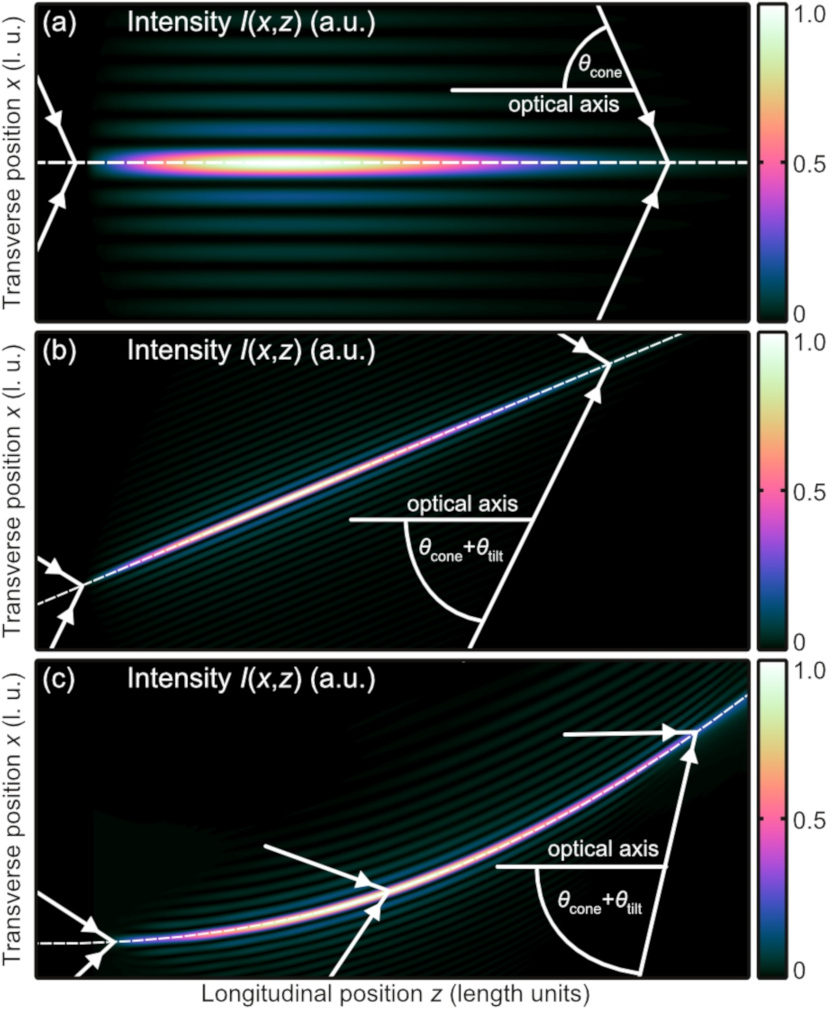}
    \caption{Simulated propagation behavior of a conventional Bessel--Gaussian beam\cite{mcgloin2005bessel} (a), a tilted Bessel--Gaussian beam\cite{jenne2018high} (b) and a Bessel-Gaussian beam propagating along an accelerating trajectory\cite{chremmos2012bessel} (c) to demonstrate limitations with respect to achievable angles for tailored-edge cleaving. In all cases, normalized intensity distributions $I\left(x,z\right)$ are depicted as well as the nondiffracting beam's cone angles $\theta_{\text{cone}}$ and local tilt angles $\theta_{\text{tilt}}$. The optical axis is parallel to the $z$ axis. Successful shaping of the depicted examples is enabled if the focusing unit's NA
    exceeds the sum of both angles $\text{NA} \gtrsim \theta_{\text{cone}} + \theta_{\text{tilt}}$. Here and in some of the following figures, we make use of Green's colormap scheme.\cite{green2011colour}}
    \label{fig:meinbess}
\end{figure}

For some time now, the structured light community\cite{rubinsztein2016roadmap, woerdemann2012structured} has been working on accelerating focus distributions where light is propagating along arbitrary curved trajectories. Probably the best-known representative is the Airy beam,\cite{siviloglou2007accelerating, baumgartl2008optically} which has already been used in 2012 by Mathis \textit{et al.}~for micromachining  different types of materials along curves.\cite{mathis2012micromachining} Such concepts have also been used in recent studies to machine glasses with C-shaped edges.\cite{sohr2021shaping, ungaro2021single} Related optical approaches were used to generate accelerating Bessel-like beams,\cite{chremmos2012bessel, ungaro2021single} which are of particular interest for single-pass cutting of transparent materials of relevant thickness for the consumer electronics industry, $d\gtrsim\unit[0.5]{mm}$. As demonstrated in Fig.\,\ref{fig:meinbess}, where cone- ($\theta_{\text{cone}}$) and tilt angles ($\theta_{\text{tilt}}$) of curved Bessel--Gaussian beams are defined, this concept, in principle, allows processing along small radii of curvatures. However, the required NA of the focusing unit would exceed $1$, if $45^{\circ}$ trajectories are to be aimed in vacuum, cf.\,Fig.\,\ref{fig:meinbess}\,(c). For processing inside the volume of the bulk material, the entrance surface's index mismatch represents the same challenge as for machining with inclined optical heads, cf.~Sec.\,\ref{sec:level1}. Efforts for high-NA focusing and corresponding short working distances and/or index matching concepts seem too high, especially from an industrial perspective. We, therefore, conclude that the class of accelerating beams (including ``second-type nondiffracting beams'' \cite{baumgartl2008optically, woerdemann2012structured}) cannot be used for tailored-edge cleaving if $\alpha \rightarrow 45^{\circ}$ and a relevant substrate thickness of $d\gtrsim\unit[0.5]{mm}$ is desired, cf.~Fig.\,\ref{fig:champion}. This is also confirmed by recent C-shaped glass cutting experiments demonstrating $\alpha>70^{\circ}$ using Airy beams.\cite{sohr2021shaping, ungaro2021single}

The concept presented here is not a beam shaping approach in the classical sense, but distributes nearly ideal focus copies along an arbitrary trajectory in a well-defined working volume of a focusing unit. Diffractive beam splitters are common to the materials processing community enabling parallel processing and throughput scaling for, e.g., drilling or ablation applications exploiting the complete power/energy performance of the laser system.\cite{kumkar2017throughput} Usually, foci are split in a single propagation plane (2D-beam splitting)\cite{wyrowski1994use}---often combined with optical scanners and $f$-$\theta$-lenses.\cite{flamm2019beam} The expansion to the third spatial dimension (propagation direction: $z$ axis), thus 3D-beam splitting, is a technique successfully used for parallel micromanipulation,\cite{zhu2014three} multifocal microscopy,\cite{zhu2014three} optical data storage,\cite{gu2014optical} and for micromachining  the volume of transparent materials (ultrafast laser writing\cite{jesacher2010parallel} and welding\cite{flamm2019beam}). The holographic approach foresees to displace a number of foci from the original geometric focus position determined by the focusing unit in transverse and longitudinal directions. For $2f$-configurations, as used in this work, the design algorithm and the experimental verification is provided in Refs.~\citenum{flamm2021structured} and \citenum{flamm2019beam}, respectively. Here, holographic tilt and defocus transmission functions\cite{valle2012analytic} are multiplexed (one for each spot to be split) with well-defined phase relations resulting in particularly efficient and homogeneous focus distributions.\cite{flamm2021structured} The final phase-only hologram can be displayed by a flexible liquid-crystal-on-silicon-based spatial light modulator (SLM) or realized by a static DOE.

Although, as already mentioned, we pursue a beam splitting concept, in the following, we will refer to the entirety of the foci as \textit{one} focus distribution. This is due to the fact that simultaneity in the introduction of the material modifications is decisive for the process presented. It is the simultaneous impact of the large number of foci to the material that generates connected modifications without shielding effects which, in turn, enables the substrate separation in the first place.

\begin{figure}
    \centering
    \includegraphics[width=0.48\textwidth]{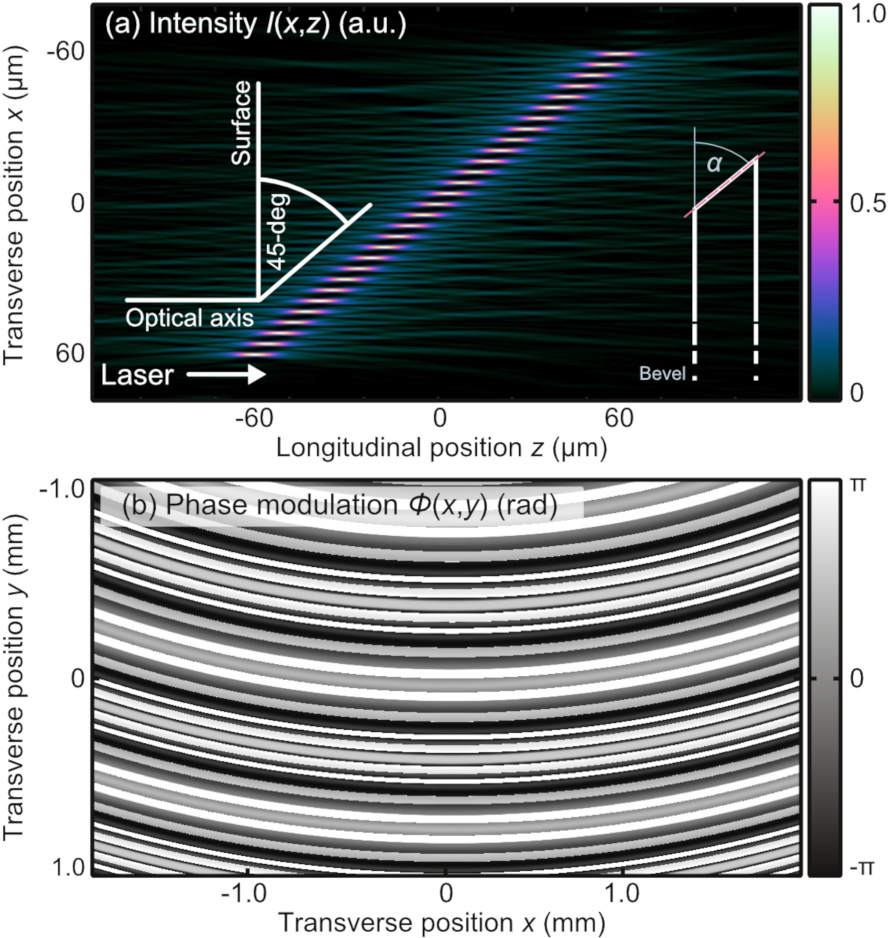}
    \caption{Example of a 3D-focus distribution for tailored-edge cleaving with simulated intensities $I\left(x,z\right)$ in an isotropic medium with $n=1.45$ at $\lambda = \unit[1030]{nm}$. The $z$ axis corresponds to light's propagation direction. Beam splitting of $29$ Gaussian foci along a 45-deg trajectory with cathetus length of $\unit[120]{\upmu m}$ (a). For illustration purposes, the arrangement of the focal points along the target geometry is shown schematically. Central detail of the corresponding phase modulation $\Phi\left(x,y\right)$ of the holographic beam splitter (b).}
    \label{fig:C1}
\end{figure}

In Fig.\,\ref{fig:C1}, an example of a 3D-focus \textcolor{black}{distribution} is depicted. Here, in Fig.\,\ref{fig:C1}\,(a), $29$ Gaussian foci are equidistantly distributed along a $45^{\circ}$ trajectory as can be seen in the intensity representation $I\left(x,z\right)$ which is normalized to unit intensity. Note the different axis scaling of longitudinal ($z$ axis) to transverse ($x$ axis) spatial dimensions. In the simulations, focusing was realized using an objective lens with $\text{NA} = 0.4$. The cathetus length of the shaped focus distribution for processing in an isotropic medium with $n=1.45$ at our processing wavelength of $\lambda = \unit[1030]{nm}$ is by design $\unit[120]{\upmu m}$. The uniformity error, defined as the maximum peak intensity difference of the $29$ spots, is a few percentage points only. The phase-only realization of the transmission function will always yield optical powers in unwanted diffraction orders, see, e.g., the weak local intensity maximum at $\left(x=\unit[65]{\upmu m}, z=\unit[65]{\upmu m}\right)$, in Fig.\,\ref{fig:C1}\,(a). Here, relative peak intensities are below $15\%$. The phase modulation $\Phi\left(x,y\right)$ of the designed phase-only transmission function $T\left(x,y\right) = \exp{\left[i\Phi\left(x,y\right)\right]}$ can be seen in Fig.\,\ref{fig:C1}\,(b) (central details). Although the number of multiplexed channels is already high, tilt and defocus functions from our design approach are apparent. Depending on the way of realizing $\Phi\left(x,y\right)$, for example, using a digital hologram or a static diffractive optical element, diffraction efficiencies may exceed $90\,\%$.\cite{flamm2021structured} Spatial separation of the spots with respect to their neighbors and prevention of unwanted interference effects, like intensity fluctuations, is achieved by choosing a sufficiently large diameter of the Gaussian raw beam. In this case, theoretically, the beam quality of the single spots is at the diffraction limit, thus $M^2\approxeq1$. As the 3D-beam splitter generates ideal focus copies, in practice, beam quality is determined by the optical field of the illuminating beam and the quality of the focusing unit.

\begin{figure}
    \centering
    \includegraphics[width=0.48\textwidth]{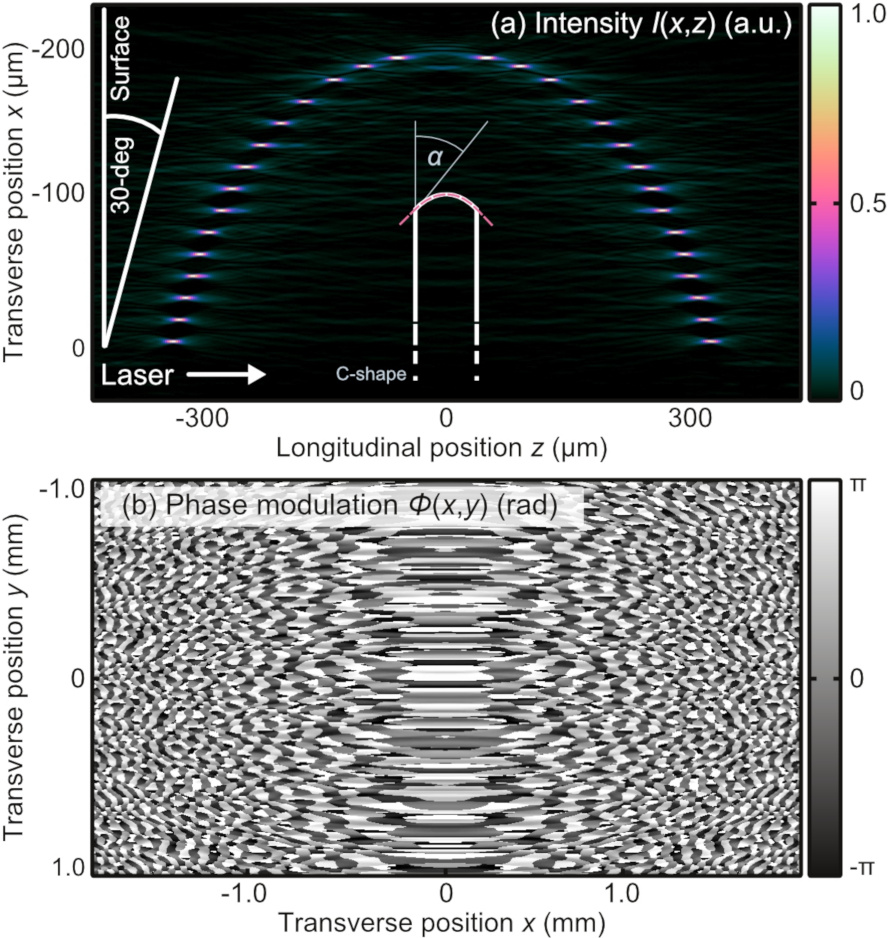}
    \caption{Example of a 3D-focus distribution for C-shape edge cleaving with simulated intensities $I\left(x,z\right)$ in an isotropic medium with $n=1.45$ at $\lambda = \unit[1030]{nm}$. The $z$ axis corresponds to light's propagation direction. Beam splitting of $30$ Gaussian foci along a C-shaped trajectory (a). For illustration purposes, the arrangement of the focal points along the target geometry is shown schematically. Central detail of the corresponding phase modulation $\Phi\left(x,y\right)$ of the holographic beam splitter (b).}
    \label{fig:C2}
\end{figure}

A second example is discussed by means of Fig.\,\ref{fig:C2} where $30$ Gaussian foci are distributed along a C-shaped trajectory. The simulated and normalized intensity profile of the focus distribution $I\left(x,z\right)$ proves high-quality 3D-beam splitting with uniformity errors below $10\,\%$, see Fig.\,\ref{fig:C2}\,(a). 
The ``length'' of the C-shaped focus distribution is adapted to induce modifications inside the volume of $\unit[550]{\upmu m}$-thick display glass. Tangential angles to this trajectory amount to $\alpha\sim\left(30^{\circ} \dots 90^{\circ}\right)$, see white angle in Fig.~\ref{fig:C2}\,(a). Note the complete different axis scaling compared to Fig.~\ref{fig:C1}\,(a). The phase modulation $\Phi\left(x,y\right)$ of the corresponding holographic beam splitter is shown in Fig.~\ref{fig:C2}\,(b). Here, even a trained eye would have difficulties predicting the optical effect of this transmission containing lens and tilt transmission functions similar to the first example, cf.~Fig.~\ref{fig:C1}\,(b). However, one can explain the present phase profile in such a way that the individual foci of the target distribution in the inverse propagation to the rear image plane of the objective form a coherent field, whose phase distribution is shown here. As required spatial frequencies are significantly increased with respect to the first example, cf.~Fig.~\ref{fig:C1}\,(b), slightly reduced maximum diffraction efficiencies ($>85\%$) are expected, again depending on the efforts for realizing the hologram.

From an optical point of view, the problem of creating high intensities along advanced edge contours seems to be solved. However, for the design of the holographic 3D-beam splitter only linear optical concepts (ideal paraxial focusing, wave optical propagation based on spectrum-of-plane-wave operators, Snell's and Fresnel's laws for the interface transition etc.) were taken into consideration. It is not revealed here in a theoretical manner what the associated modifications look like and whether they are suitable for a separation step. Although the focal distributions, i.e.~the entirety of the beam-split foci, may be arranged along accelerated trajectories, the individual Gaussian foci still propagate parallel to the optical axis, cf.~Figs.~\ref{fig:C1} and \ref{fig:C2}. At first glance, this situation is unfavorable for material processing, since we expect elongated modifications aligned along this axis,\cite{grossmann2016transverse, bergner2018spatio} which, depending on the edge contour, will extend into the useful part of the workpiece. We want to clarify this notoriously complex situation experimentally by analyzing the laser-induced modifications in the next section and by applying a selective etching strategy for the actual separation.

\section{\label{sec:Expo}Single-pass tailored edge cleaving}
\begin{figure*}[]
    \centering
    \includegraphics[width=.95\textwidth]{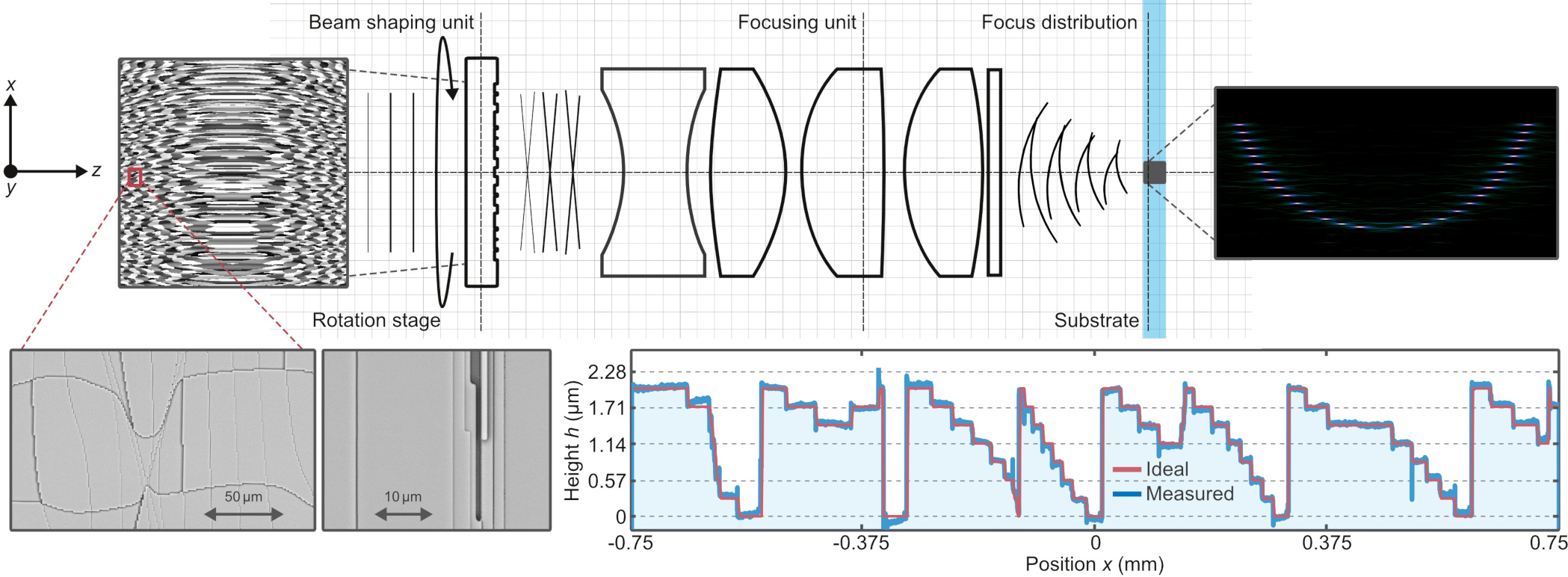}
        \caption{Optical setup for single-pass tailored-edge cleaving of transparent materials. The central beam shaping element is illuminated by free-space or fiber-guided ultrashort pulses emerged from TruMicro Series lasers.\cite{jansen2018pulsed} Phase-only volume beam splitting is realized by flexible liquid-crystal-on-silicon-based spatial light modulators or by stationary diffractive optical elements. Modulated light is focused using adapted long-working distance and large working volume, high-power suitable microscope objectives.\cite{flamm2019beam} During processing, the workpiece is translated relative to the optical head in the $y$ direction, see coordinate system. For machining along contours, a rotational stage is required that allows us to rotate the central beam shaping element or the entire processing optics around the optical axis ($z$ axis). In any case, processing is performed under vertical illumination. Inclination of the optical head\cite{jenne2018high, ungaro2021using, marjanovic2019edge} is not required. }
    \label{fig:setup}
\end{figure*}

The optical concepts from Sec.~\ref{sec:Depo} are applied to generate C-shaped focus distributions for tailored-edge cleaving of nonstrengthened Corning Gorilla® glass substrates. Ultrashort laser pulses emerged from a TruMicro Series laser\cite{jansen2018pulsed} are illuminating a central beam shaping element acting as holographic phase-only 3D-beam splitter in a $2f$-configuration, see Fig.~\ref{fig:setup}.\cite{flamm2021structured}
We conducted experiments with flexible liquid-crystal-on-silicon-based SLM (Hamamatsu X13139-03) and stationary DOEs. The latter were realized as quantized eight-level elements fabricated via laser lithography in fused silica exhibiting efficiencies of $\sim\unit[90]{\%}$. Different microscope images of this element are depicted at the bottom of Fig.\,\ref{fig:setup} confirming successful writing and etching steps with minor height profile differences and resulting neglectable zeroth-order power of $<\unit[0.5]{\%}$. The design of the 3D-beam splitter is already provided in Fig.\,\ref{fig:C2} and is adapted to the substrate's thickness of $d=\unit[550]{\upmu m}$, cf.~Fig.\,\ref{fig:champion}. When working with the flexible SLM, unmodulated light is implemented into our focus distribution and appears as zeroth-order at $\left(x=0,\, z=0\right)$, cf.~Fig.\,\ref{fig:C2}\,(a), in the geometric focus of the objective. Focusing is realized by a long-working distance microscope objective ($\text{NA} = 0.4$, $f=\unit[10]{mm}$) optimized for the operational wavelength at $\lambda = \unit[1030]{nm}$ and the present working volume. Machining is achieved during translating the workpiece relative to the optical head in $y$ direction, see coordinate system in Fig.\,\ref{fig:setup}. For contour cutting, the orientation of the shaped focus distribution needs to be rotated around the $z$ axis which can be realized by either rotating the entire optical head or the central beam shaping element. In the latter case, machining with the DOE is preferred resulting in a particularly compact and robust processing optics, cf.~Fig.\,\ref{fig:setup}.

\begin{figure}
    \centering
    \includegraphics[width=0.48\textwidth]{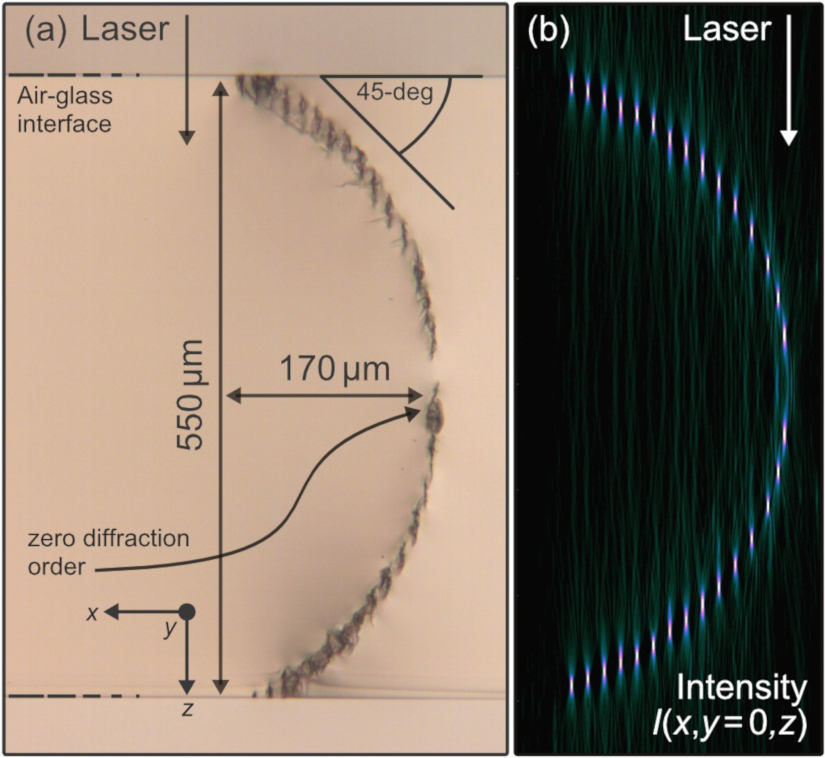}
    \caption{Microscope image of the edge of an unhardened $\unit[550]{\upmu m}$-thick Corning Gorilla® glass substrate with laser-induced C-shaped modifications (a). Spatially separated  type-III-regime modifications\cite{itoh2006ultrafast} caused by the split Gaussian foci are apparent. Tangential angles to the modified contour and surface are reduced down to $\alpha \lesssim 30^{\circ}$. Processing was achieved parallel to the $y$ axis, see the coordinate system. For reasons of comparability, we again depict the intensity profile of the corresponding focus distribution with adapted axis scaling (b), cf.\,Fig.\,\ref{fig:C2}.}
    \label{fig:C-mik}
\end{figure}

The result of the laser modification process is shown in Fig.\,\ref{fig:C-mik}. Here, a microscope image recorded perpendicularly from the edge of an unhardened $\unit[550]{\upmu m}$-thick Corning Gorilla® glass substrate can be seen, see Fig.\,\ref{fig:C-mik}\,(a). While focusing was achieved under vertical incidence parallel to the $z$ axis, see the coordinate system, the workpiece was translated in the $y$ direction with respect to the optical head. The holographically split $31$ Gaussian foci (including the zero diffraction order) cause spatially separated modifications, which clearly follow the desired C-shaped contour. We can, therefore, answer the question posed at the beginning of Sec.\,\ref{sec:Depo} with a resounding Yes: Holographic 3D-beam splitters enable us to deposit energy along arbitrary trajectories, including C-shaped edges with $\alpha < 45^{\circ}$.

During our modification process, laser parameters were chosen to mainly generate type-III-regime modifications \cite{itoh2006ultrafast} inside the glass volume. Here, a pulse energy of $\lesssim \unit[300]{\upmu J}$ was equally distributed to picosecond pulse trains\cite{herman2003burst, kerse2016ablation} emerging from a TruMicro 2000 Series laser.\cite{jansen2018pulsed} Feed rates were chosen to generate a modification pitch of $\sim\unit[5]{\upmu m}$ ( the $y$ direction, cf.~Fig.\,\ref{fig:setup}). Please note, that the stated laser parameters of this study represent useful values and can form the basis for future investigations. However, we do not claim to have found the optimum ones. Depending on substrate geometry and material, adapted laser parameters need to be found. In addition, there will be a strong dependency on which the separation process (for example, chemical versus thermal) is actually aimed at.\cite{rave2021glass} \textcolor{black}{Particularly when it comes to thermal separation with CO$_2$-laser radiation---highly relevant from an industrial point of view---the required laser parameters will differ completely and an additional parameter study would be needed.}

As each of the $31$ foci propagates parallel to the $z$ axis, cf.~Fig.\,\ref{fig:C2}\,(a), the corresponding modifications exhibit a clear elongation along this direction---a well-known behavior when using Gaussian focus distributions for processing alkali-aluminosilicate glasses.\cite{grossmann2016transverse, bergner2018spatio,jenne2018multi} From the microscope image [Fig.\,\ref{fig:C-mik}\,(a)], the modification dimensions along the $z$ axis amount to $\unit[\left(20\dots30\right)]{\upmu m}$, which is about twice the Rayleigh length of our foci. Although these modifications protrude into the useful part of the workpiece, especially on the top and bottom sides, our goal is to remove an edge contour in C-shape. 

For this second processing step, various strategies are known, for example, by applying mechanical loads\cite{jenne2020facilitated} or by inducing thermal stresses from CO$_2$-laser radiation.\cite{nisar2013laser, ungaro2021using} It is also well-studied that different types of laser-induced modifications can exhibit much larger etch rates than the untreated glass volume ($>1000:1$).\cite{gottmann2017selective} This selective laser-induced etching concept enables rapid fabrication of 3D-glass structures of arbitrary shape with smallest structural features down to the $\unit[10]{\upmu m}$-scale.\cite{hermans2014selective, gottmann2017selective} We would like to emphasize that the actual separation concept will require adapted laser parameters and focus properties.

\begin{figure}
    \centering
    \includegraphics[width=0.48\textwidth]{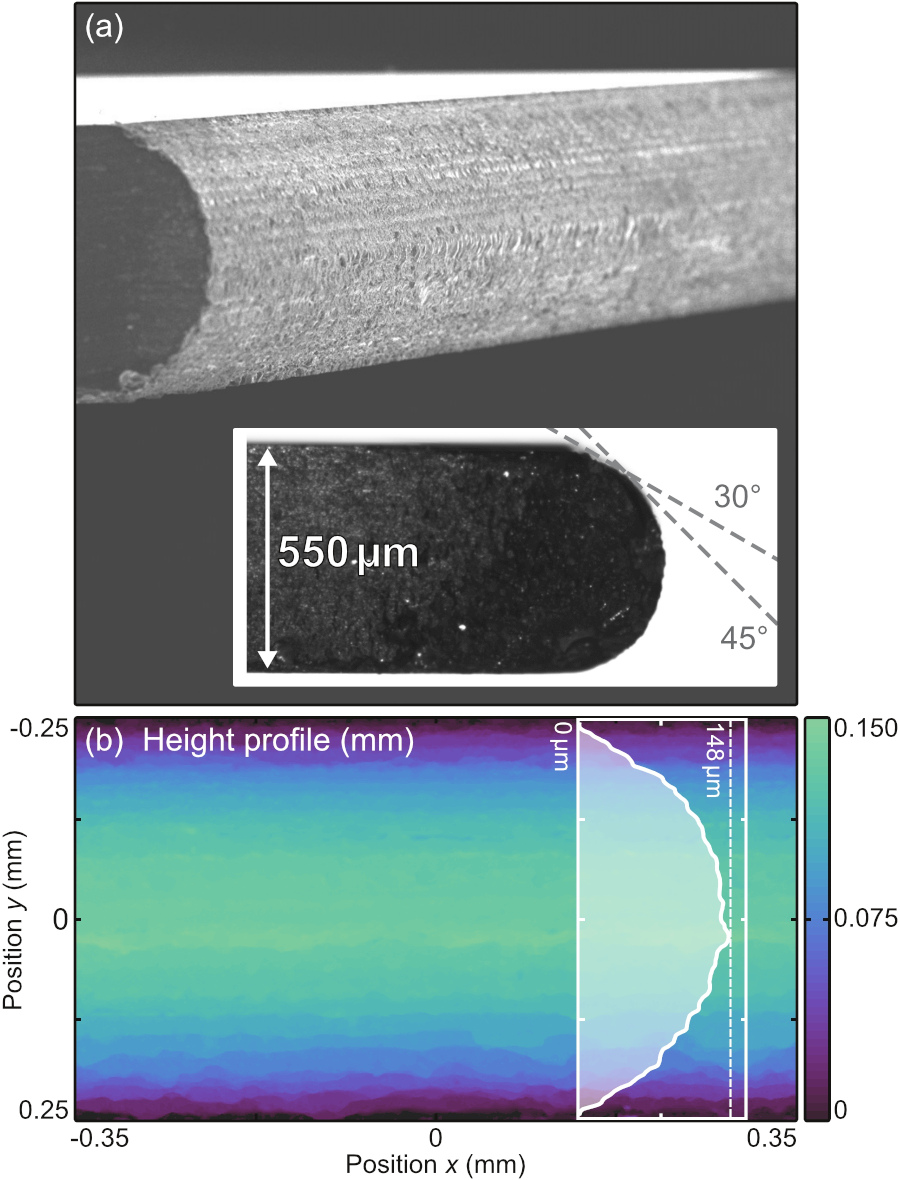}
    \caption{Various microscope images confirming the successful processing of the glass substrate with an edge contour in C-shape. Scanning electron image (a) with a light microscope recording depicted in the inset. The tangential angles to the modified contour prove $\alpha \lesssim 30^{\circ}$, cf.\,Fig.\,\ref{fig:champion}. The reconstructed height profile recorded perpendicular to the C-shaped surface with a laser scanning microscope is shown in (b).}
    \label{fig:SEM}
\end{figure}

Our SLE strategy is based on applying $30$\,wt.-\% KOH solution to the laser-modified substrate in an ultrasonic bath operating at $>\,80\,^{\circ}$C.\cite{kaiser2019selective, rave2021glass} After an etching period of $<\unit[60]{min}$ separation is achieved. The processing result is depicted in Fig.\,\ref{fig:SEM} where different microscope images prove a successful tailored-edge processing. With the SEM image in Fig.\,\ref{fig:SEM}\,(a) a perspective impression of the etched workpiece is provided.
In the inset, a light microscope recording the profile of the shaped edge [parallel to the processing direction ($y$ direction), cf.\,Fig.\,\ref{fig:C-mik}] can be seen proving that tangential angles to the surface are reduced to $\alpha \lesssim 30^{\circ}$, cf.\,Fig.\,\ref{fig:champion}. The result of a surface profile measurement, now, recorded perpendicular to the C-shaped surface (parallel to the $x$ direction, cf.\,Fig.\,\ref{fig:C-mik}) with a laser scanning microscope is depicted in Fig.\,\ref{fig:SEM}\,(b).

\begin{figure*}
    \centering
    \includegraphics[width=0.8\textwidth]{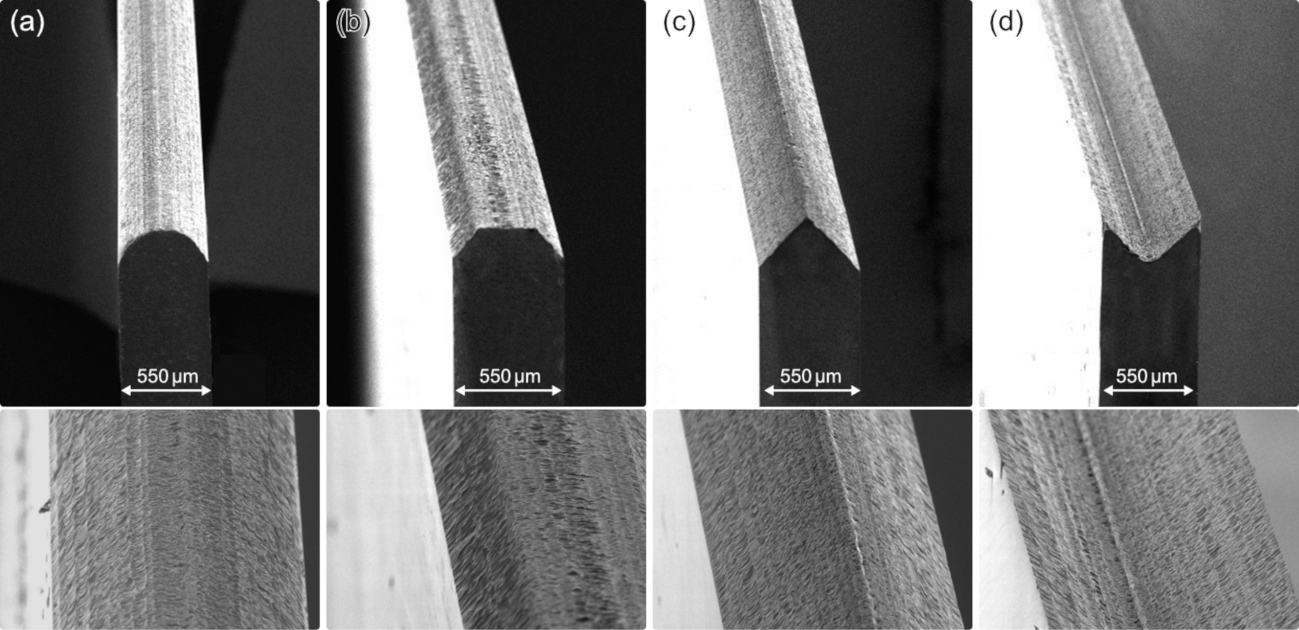}
    \caption{Menu of glass substrates with tailored-edges investigated with scanning electron microscopy. (a) C-shaped and (b) chamfered edge, cf.\,Fig.\,\ref{fig:champion}. Examples shown in (c) and (d) ($90^{\circ}$ apex profile and inverted $90^{\circ}$ apex profile) do not fulfill the motivation of a protected edge, cf.\,Sec.\,\ref{sec:level1}, but are nevertheless shown here to demonstrate the versatility of our concept. Detailed edge images are assigned to all cases in the bottom row. In all cases, $\unit[550]{\upmu m}$-thick Corning Gorilla® glass was processed with a single laser pass and subsequent chemical etching.}
    \label{fig:pfu}
\end{figure*}

Although we are only taking a closer look at one edge shape here, the concept presented allows an enormous variety of geometries. Figure \ref{fig:pfu} provides a menu of different edge contours---all laser modified within a single pass.

\section{Mechanical properties of edge-chamfered substrates}\label{sec:mech}
Microscopic inspection of the glass edge (Fig.~\ref{fig:pfu}) or determination of the edge roughness (Fig.~\ref{fig:SEM}) provide important process information, but do not confirm enhanced mechanical properties, cf.~Sec.~\ref{sec:level1}. Following the argumentation of Marjanovic \textit{et al.},\cite{marjanovic2019edge} cf.~Sec.~\ref{sec:level1}, we are convinced that tailored glass edges with reduced tangential angles to the surface, see Fig.\,\ref{fig:champion}\,(b)\,--\,(d), do not exhibit increased edge strength \textit{per se}, but show enhanced mechanical properties in case of impacts or material defects. The edge chamfering method---in our case the combination of inducing laser modifications and chemical separation, cf.\,Sec.~\ref{sec:Expo}---must, therefore, not reduce the substrate's bending strength. In this first, brief investigation, we are content to verify this and refer to later studies that will test for impact strength, in particular, also for hardened samples. 

In the following, bending strength is quantified by determining the maximal flexural load applied to C-shaped tailored-edge samples (Corning Gorilla®, nonstrengthened, $\unit[550]{\upmu m}$ thickness) via four-point bending tests.\cite{bukieda2020study} Conventionally laser-cut samples of equal geometry and material but with straight face edges, cf.~Fig.\,\ref{fig:champion}\,(a), serve as reference. Here, separation was achieved mechanically. These reference substrates were cut to size with nondiffracting beams from a TOP Cleave cutting optics\cite{flamm2021ultrafast} and picosecond pulse trains emerging from a TruMicro 2000 Series laser.\cite{jansen2018pulsed}
\begin{figure}
    \centering
    \includegraphics[width=.48\textwidth]{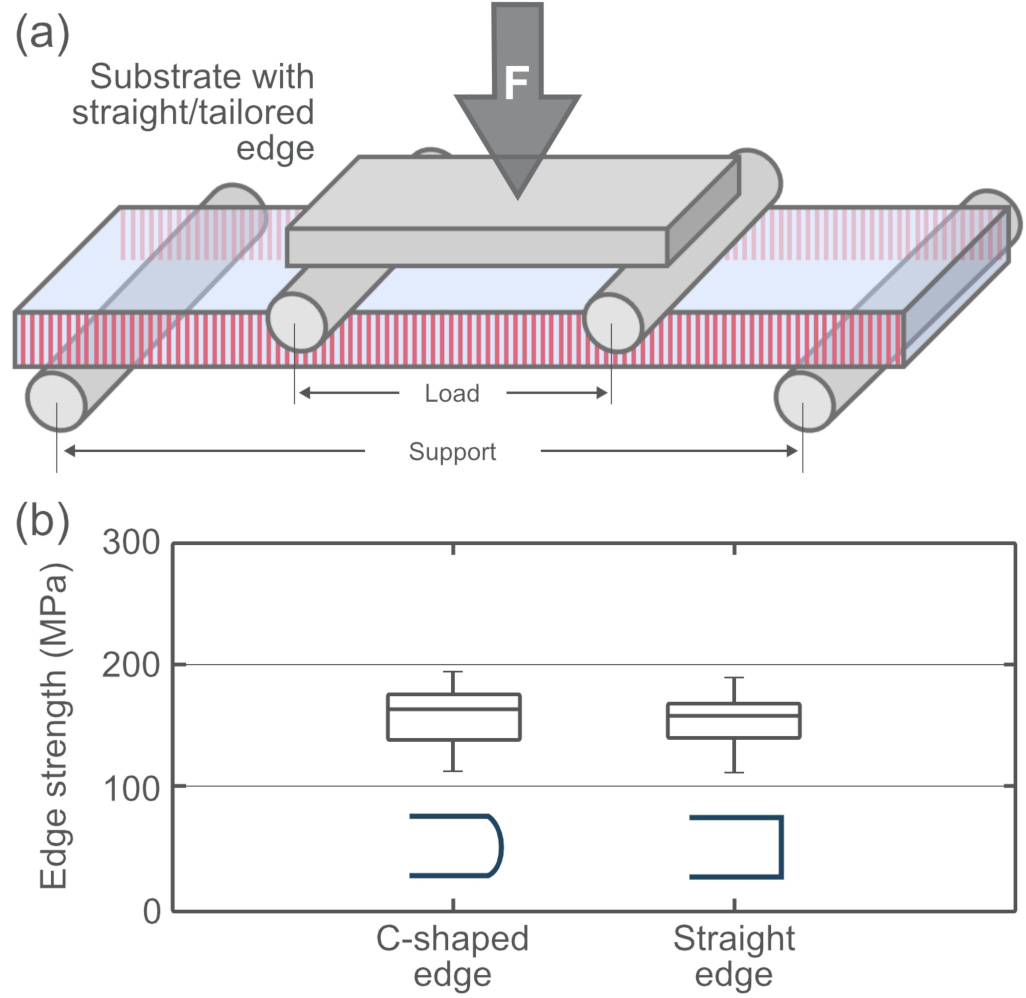}
    \caption{Schematic of the four-point bending test arrangement (a) with definition of pin orientation and substrate edge. Result of bending tests in box-plot representation (b) with measured flexural loads of C-shaped vs.~straight substrate edges.}
    \label{fig:4PBT}
\end{figure}

A schematic of the bending test arrangement is depicted in Fig.~\ref{fig:4PBT}\,(a) where the orientation of the loading pins to the substrate's processed edge is defined. The ratio of load span to support span during our experiments was set to $0.5$. Results of the bending tests for both edge geometries are box-plotted in Fig.~\ref{fig:4PBT}\,(b). Here, the edge strength median of the C-shaped substrates was determined to be $\sim\unit[160]{MPa}$ and, thus, even slightly exceeds that of the straight substrates by $3$\,\%. Attaching the shaped edge via inducing laser modifications and applying chemical etching, thus, protects the glass article and does not impair bending properties which are even slightly improved.

\textcolor{black}{We would like to point out once again that only a first, brief study could be provided here already indicating enhanced mechanical properties. Forthcoming, more detailed studies have to provide bending tests for strengthened substrates separated also via CO$_2$ laser radiation. Ideally, the reference samples were separated with the same strategy as the tailored-edge cleaved substrates (for example, chemical versus chemical separation).}

\section{Conclusion}
We reported on the cleaving of glass substrates with various edge contours including C-shapes using ultrashort laser pulses and a 3D holographic beam splitting concept. The single-pass, full thickness laser modification step enables to generate tailored edges with tangential angles to the surface of $45^{\circ}$---world's first, to the best of our knowledge. Using TruMicro Series ultrafast laser platforms feed rates up to $\unit[]{m/s}$ become possible. In this study, laser modified $\unit[550]{\upmu m}$-thick glass samples were separated by chemical etching within less than $\unit[1]{h}$. Other separation strategies are conceivable. The mechanical properties of processed glass articles were investigated with four-point-bending tests reaching bend resistances that even exceed those of substrates with straight edges. 

\section*{Acknowledgments}
The authors acknowledge the support of Hamza Dounassre (TRUMPF Laser- und Systemtechnik GmbH) who performed the bending tests presented in Fig.~\ref{fig:4PBT} and Wan Nur Syahirah Binti Mokhtar (TRUMPF Laser- und Systemtechnik GmbH) for assisting with the simulations depicted in Fig.~\ref{fig:C2}.

\nocite{*}

\bibliography{02_Litbib}

\providecommand{\noopsort}[1]{}\providecommand{\singleletter}[1]{#1}%
\begin{thebibliography}{37}%
\makeatletter
\providecommand \@ifxundefined [1]{%
 \@ifx{#1\undefined}
}%
\providecommand \@ifnum [1]{%
 \ifnum #1\expandafter \@firstoftwo
 \else \expandafter \@secondoftwo
 \fi
}%
\providecommand \@ifx [1]{%
 \ifx #1\expandafter \@firstoftwo
 \else \expandafter \@secondoftwo
 \fi
}%
\providecommand \natexlab [1]{#1}%
\providecommand \enquote  [1]{``#1''}%
\providecommand \bibnamefont  [1]{#1}%
\providecommand \bibfnamefont [1]{#1}%
\providecommand \citenamefont [1]{#1}%
\providecommand \href@noop [0]{\@secondoftwo}%
\providecommand \href [0]{\begingroup \@sanitize@url \@href}%
\providecommand \@href[1]{\@@startlink{#1}\@@href}%
\providecommand \@@href[1]{\endgroup#1\@@endlink}%
\providecommand \@sanitize@url [0]{\catcode `\\12\catcode `\$12\catcode
  `\&12\catcode `\#12\catcode `\^12\catcode `\_12\catcode `\%12\relax}%
\providecommand \@@startlink[1]{}%
\providecommand \@@endlink[0]{}%
\providecommand \url  [0]{\begingroup\@sanitize@url \@url }%
\providecommand \@url [1]{\endgroup\@href {#1}{\urlprefix }}%
\providecommand \urlprefix  [0]{URL }%
\providecommand \Eprint [0]{\href }%
\providecommand \doibase [0]{http://dx.doi.org/}%
\providecommand \selectlanguage [0]{\@gobble}%
\providecommand \bibinfo  [0]{\@secondoftwo}%
\providecommand \bibfield  [0]{\@secondoftwo}%
\providecommand \translation [1]{[#1]}%
\providecommand \BibitemOpen [0]{}%
\providecommand \bibitemStop [0]{}%
\providecommand \bibitemNoStop [0]{.\EOS\space}%
\providecommand \EOS [0]{\spacefactor3000\relax}%
\providecommand \BibitemShut  [1]{\csname bibitem#1\endcsname}%
\let\auto@bib@innerbib\@empty
\bibitem [{\citenamefont {Jenne}\ \emph {et~al.}(2020)\citenamefont {Jenne},
  \citenamefont {Flamm}, \citenamefont {Chen}, \citenamefont {Sch{\"a}fer},
  \citenamefont {Kumkar},\ and\ \citenamefont {Nolte}}]{jenne2020facilitated}%
  \BibitemOpen
  \bibfield  {author} {\bibinfo {author} {\bibfnamefont {M.}~\bibnamefont
  {Jenne}}, \bibinfo {author} {\bibfnamefont {D.}~\bibnamefont {Flamm}},
  \bibinfo {author} {\bibfnamefont {K.}~\bibnamefont {Chen}}, \bibinfo {author}
  {\bibfnamefont {M.}~\bibnamefont {Sch{\"a}fer}}, \bibinfo {author}
  {\bibfnamefont {M.}~\bibnamefont {Kumkar}}, \ and\ \bibinfo {author}
  {\bibfnamefont {S.}~\bibnamefont {Nolte}},\ }\bibfield  {title} {\enquote
  {\bibinfo {title} {Facilitated glass separation by asymmetric {Bessel-like}
  beams},}\ }\href@noop {} {\bibfield  {journal} {\bibinfo  {journal} {Optics
  Express}\ }\textbf {\bibinfo {volume} {28}},\ \bibinfo {pages} {6552--6564}
  (\bibinfo {year} {2020})}\BibitemShut {NoStop}%
\bibitem [{\citenamefont {Marjanovic}\ \emph {et~al.}(2019)\citenamefont
  {Marjanovic}, \citenamefont {Andrew}, \citenamefont {Garrett}, \citenamefont
  {Piech}, \citenamefont {Quintal}, \citenamefont {Schillinger}, \citenamefont
  {Tsuda}, \citenamefont {Wagner},\ and\ \citenamefont
  {Yeary}}]{marjanovic2019edge}%
  \BibitemOpen
  \bibfield  {author} {\bibinfo {author} {\bibfnamefont {S.}~\bibnamefont
  {Marjanovic}}, \bibinfo {author} {\bibfnamefont {D.}~\bibnamefont {Andrew}},
  \bibinfo {author} {\bibfnamefont {P.}~\bibnamefont {Garrett}}, \bibinfo
  {author} {\bibfnamefont {A.}~\bibnamefont {Piech}}, \bibinfo {author}
  {\bibfnamefont {J.~M.}\ \bibnamefont {Quintal}}, \bibinfo {author}
  {\bibfnamefont {H.}~\bibnamefont {Schillinger}}, \bibinfo {author}
  {\bibfnamefont {S.}~\bibnamefont {Tsuda}}, \bibinfo {author} {\bibfnamefont
  {R.~S.}\ \bibnamefont {Wagner}}, \ and\ \bibinfo {author} {\bibfnamefont
  {A.~N.}\ \bibnamefont {Yeary}},\ }\href@noop {} {\enquote {\bibinfo {title}
  {Edge chamfering methods},}\ } (\bibinfo {year} {Oct. 15 2019}),\ \bibinfo
  {note} {{US} Patent 10,442,719}\BibitemShut {NoStop}%
\bibitem [{\citenamefont {Bukieda}\ \emph {et~al.}(2020)\citenamefont
  {Bukieda}, \citenamefont {Lohr}, \citenamefont {Meiberg},\ and\ \citenamefont
  {Weller}}]{bukieda2020study}%
  \BibitemOpen
  \bibfield  {author} {\bibinfo {author} {\bibfnamefont {P.}~\bibnamefont
  {Bukieda}}, \bibinfo {author} {\bibfnamefont {K.}~\bibnamefont {Lohr}},
  \bibinfo {author} {\bibfnamefont {J.}~\bibnamefont {Meiberg}}, \ and\
  \bibinfo {author} {\bibfnamefont {B.}~\bibnamefont {Weller}},\ }\bibfield
  {title} {\enquote {\bibinfo {title} {Study on the optical quality and
  strength of glass edges after the grinding and polishing process},}\
  }\href@noop {} {\bibfield  {journal} {\bibinfo  {journal} {Glass Structures
  \& Engineering}\ }\textbf {\bibinfo {volume} {5}},\ \bibinfo {pages}
  {411--428} (\bibinfo {year} {2020})}\BibitemShut {NoStop}%
\bibitem [{\citenamefont {Jenne}\ \emph
  {et~al.}(2018{\natexlab{a}})\citenamefont {Jenne}, \citenamefont {Flamm},
  \citenamefont {Ouaj}, \citenamefont {Hellstern}, \citenamefont {Kleiner},
  \citenamefont {Grossmann}, \citenamefont {Koschig}, \citenamefont {Kaiser},
  \citenamefont {Kumkar},\ and\ \citenamefont {Nolte}}]{jenne2018high}%
  \BibitemOpen
  \bibfield  {author} {\bibinfo {author} {\bibfnamefont {M.}~\bibnamefont
  {Jenne}}, \bibinfo {author} {\bibfnamefont {D.}~\bibnamefont {Flamm}},
  \bibinfo {author} {\bibfnamefont {T.}~\bibnamefont {Ouaj}}, \bibinfo {author}
  {\bibfnamefont {J.}~\bibnamefont {Hellstern}}, \bibinfo {author}
  {\bibfnamefont {J.}~\bibnamefont {Kleiner}}, \bibinfo {author} {\bibfnamefont
  {D.}~\bibnamefont {Grossmann}}, \bibinfo {author} {\bibfnamefont
  {M.}~\bibnamefont {Koschig}}, \bibinfo {author} {\bibfnamefont
  {M.}~\bibnamefont {Kaiser}}, \bibinfo {author} {\bibfnamefont
  {M.}~\bibnamefont {Kumkar}}, \ and\ \bibinfo {author} {\bibfnamefont
  {S.}~\bibnamefont {Nolte}},\ }\bibfield  {title} {\enquote {\bibinfo {title}
  {High-quality tailored-edge cleaving using aberration-corrected {Bessel-like}
  beams},}\ }\href@noop {} {\bibfield  {journal} {\bibinfo  {journal} {Optics
  Letters}\ }\textbf {\bibinfo {volume} {43}},\ \bibinfo {pages} {3164--3167}
  (\bibinfo {year} {2018}{\natexlab{a}})}\BibitemShut {NoStop}%
\bibitem [{\citenamefont {Ungaro}\ \emph {et~al.}(2021)\citenamefont {Ungaro},
  \citenamefont {Kaliteevskiy}, \citenamefont {Sterlingov}, \citenamefont
  {Ivanov}, \citenamefont {Ruffin}, \citenamefont {Terbrueggen},\ and\
  \citenamefont {Savidis}}]{ungaro2021using}%
  \BibitemOpen
  \bibfield  {author} {\bibinfo {author} {\bibfnamefont {C.}~\bibnamefont
  {Ungaro}}, \bibinfo {author} {\bibfnamefont {N.}~\bibnamefont
  {Kaliteevskiy}}, \bibinfo {author} {\bibfnamefont {P.}~\bibnamefont
  {Sterlingov}}, \bibinfo {author} {\bibfnamefont {V.~V.}\ \bibnamefont
  {Ivanov}}, \bibinfo {author} {\bibfnamefont {A.~B.}\ \bibnamefont {Ruffin}},
  \bibinfo {author} {\bibfnamefont {R.~J.}\ \bibnamefont {Terbrueggen}}, \ and\
  \bibinfo {author} {\bibfnamefont {N.}~\bibnamefont {Savidis}},\ }\bibfield
  {title} {\enquote {\bibinfo {title} {Using phase-corrected {Bessel beams} to
  cut glass substrates with a chamfered edge},}\ }\href@noop {} {\bibfield
  {journal} {\bibinfo  {journal} {Applied Optics}\ }\textbf {\bibinfo {volume}
  {60}},\ \bibinfo {pages} {714--719} (\bibinfo {year} {2021})}\BibitemShut
  {NoStop}%
\bibitem [{\citenamefont {Flamm}\ \emph
  {et~al.}(2021{\natexlab{a}})\citenamefont {Flamm}, \citenamefont {Kleiner},
  \citenamefont {Kaiser}, \citenamefont {Zimmermann}, \citenamefont
  {Gro{\ss}mann},\ and\ \citenamefont {Kumkar}}]{flamm2021ultrafast}%
  \BibitemOpen
  \bibfield  {author} {\bibinfo {author} {\bibfnamefont {D.}~\bibnamefont
  {Flamm}}, \bibinfo {author} {\bibfnamefont {J.}~\bibnamefont {Kleiner}},
  \bibinfo {author} {\bibfnamefont {M.}~\bibnamefont {Kaiser}}, \bibinfo
  {author} {\bibfnamefont {F.}~\bibnamefont {Zimmermann}}, \bibinfo {author}
  {\bibfnamefont {D.~G.}\ \bibnamefont {Gro{\ss}mann}}, \ and\ \bibinfo
  {author} {\bibfnamefont {M.}~\bibnamefont {Kumkar}},\ }\bibfield  {title}
  {\enquote {\bibinfo {title} {Ultrafast laser cutting of transparent
  materials: the trend towards tailored edges and curved surfaces},}\ }in\
  \href@noop {} {\emph {\bibinfo {booktitle} {Laser-based Micro-and
  Nanoprocessing XV}}},\ Vol.\ \bibinfo {volume} {11674}\ (\bibinfo
  {organization} {International Society for Optics and Photonics},\ \bibinfo
  {year} {2021})\ p.\ \bibinfo {pages} {116740J}\BibitemShut {NoStop}%
\bibitem [{\citenamefont {Cheng}\ \emph {et~al.}(2020)\citenamefont {Cheng},
  \citenamefont {Xia}, \citenamefont {Kuebler},\ and\ \citenamefont
  {Yu}}]{cheng2020aberration}%
  \BibitemOpen
  \bibfield  {author} {\bibinfo {author} {\bibfnamefont {H.}~\bibnamefont
  {Cheng}}, \bibinfo {author} {\bibfnamefont {C.}~\bibnamefont {Xia}}, \bibinfo
  {author} {\bibfnamefont {S.~M.}\ \bibnamefont {Kuebler}}, \ and\ \bibinfo
  {author} {\bibfnamefont {X.}~\bibnamefont {Yu}},\ }\bibfield  {title}
  {\enquote {\bibinfo {title} {Aberration correction for {SLM-generated Bessel}
  beams propagating through tilted interfaces},}\ }\href@noop {} {\bibfield
  {journal} {\bibinfo  {journal} {Optics Communications}\ }\textbf {\bibinfo
  {volume} {475}},\ \bibinfo {pages} {126213} (\bibinfo {year}
  {2020})}\BibitemShut {NoStop}%
\bibitem [{\citenamefont {McGloin}\ and\ \citenamefont
  {Dholakia}(2005)}]{mcgloin2005bessel}%
  \BibitemOpen
  \bibfield  {author} {\bibinfo {author} {\bibfnamefont {D.}~\bibnamefont
  {McGloin}}\ and\ \bibinfo {author} {\bibfnamefont {K.}~\bibnamefont
  {Dholakia}},\ }\bibfield  {title} {\enquote {\bibinfo {title} {Bessel beams:
  diffraction in a new light},}\ }\href@noop {} {\bibfield  {journal} {\bibinfo
   {journal} {Contemporary Physics}\ }\textbf {\bibinfo {volume} {46}},\
  \bibinfo {pages} {15--28} (\bibinfo {year} {2005})}\BibitemShut {NoStop}%
\bibitem [{\citenamefont {Flamm}\ \emph
  {et~al.}(2021{\natexlab{b}})\citenamefont {Flamm}, \citenamefont {Grossmann},
  \citenamefont {Sailer}, \citenamefont {Kaiser}, \citenamefont {Zimmermann},
  \citenamefont {Chen}, \citenamefont {Jenne}, \citenamefont {Kleiner},
  \citenamefont {Hellstern}, \citenamefont {Tillkorn} \emph
  {et~al.}}]{flamm2021structured}%
  \BibitemOpen
  \bibfield  {author} {\bibinfo {author} {\bibfnamefont {D.}~\bibnamefont
  {Flamm}}, \bibinfo {author} {\bibfnamefont {D.~G.}\ \bibnamefont
  {Grossmann}}, \bibinfo {author} {\bibfnamefont {M.}~\bibnamefont {Sailer}},
  \bibinfo {author} {\bibfnamefont {M.}~\bibnamefont {Kaiser}}, \bibinfo
  {author} {\bibfnamefont {F.}~\bibnamefont {Zimmermann}}, \bibinfo {author}
  {\bibfnamefont {K.}~\bibnamefont {Chen}}, \bibinfo {author} {\bibfnamefont
  {M.}~\bibnamefont {Jenne}}, \bibinfo {author} {\bibfnamefont
  {J.}~\bibnamefont {Kleiner}}, \bibinfo {author} {\bibfnamefont
  {J.}~\bibnamefont {Hellstern}}, \bibinfo {author} {\bibfnamefont
  {C.}~\bibnamefont {Tillkorn}},  \emph {et~al.},\ }\bibfield  {title}
  {\enquote {\bibinfo {title} {Structured light for ultrafast laser micro-and
  nanoprocessing},}\ }\href@noop {} {\bibfield  {journal} {\bibinfo  {journal}
  {Optical Engineering}\ }\textbf {\bibinfo {volume} {60}},\ \bibinfo {pages}
  {025105} (\bibinfo {year} {2021}{\natexlab{b}})}\BibitemShut {NoStop}%
\bibitem [{\citenamefont {Itoh}\ \emph {et~al.}(2006)\citenamefont {Itoh},
  \citenamefont {Watanabe}, \citenamefont {Nolte},\ and\ \citenamefont
  {Schaffer}}]{itoh2006ultrafast}%
  \BibitemOpen
  \bibfield  {author} {\bibinfo {author} {\bibfnamefont {K.}~\bibnamefont
  {Itoh}}, \bibinfo {author} {\bibfnamefont {W.}~\bibnamefont {Watanabe}},
  \bibinfo {author} {\bibfnamefont {S.}~\bibnamefont {Nolte}}, \ and\ \bibinfo
  {author} {\bibfnamefont {C.~B.}\ \bibnamefont {Schaffer}},\ }\bibfield
  {title} {\enquote {\bibinfo {title} {Ultrafast processes for bulk
  modification of transparent materials},}\ }\href@noop {} {\bibfield
  {journal} {\bibinfo  {journal} {MRS Bulletin}\ }\textbf {\bibinfo {volume}
  {31}},\ \bibinfo {pages} {620--625} (\bibinfo {year} {2006})}\BibitemShut
  {NoStop}%
\bibitem [{\citenamefont {Kaiser}\ \emph {et~al.}(2019)\citenamefont {Kaiser},
  \citenamefont {Kumkar}, \citenamefont {Leute}, \citenamefont {Schmauch},
  \citenamefont {Priester}, \citenamefont {Kleiner}, \citenamefont {Jenne},
  \citenamefont {Flamm},\ and\ \citenamefont
  {Zimmermann}}]{kaiser2019selective}%
  \BibitemOpen
  \bibfield  {author} {\bibinfo {author} {\bibfnamefont {M.}~\bibnamefont
  {Kaiser}}, \bibinfo {author} {\bibfnamefont {M.}~\bibnamefont {Kumkar}},
  \bibinfo {author} {\bibfnamefont {R.}~\bibnamefont {Leute}}, \bibinfo
  {author} {\bibfnamefont {J.}~\bibnamefont {Schmauch}}, \bibinfo {author}
  {\bibfnamefont {R.}~\bibnamefont {Priester}}, \bibinfo {author}
  {\bibfnamefont {J.}~\bibnamefont {Kleiner}}, \bibinfo {author} {\bibfnamefont
  {M.}~\bibnamefont {Jenne}}, \bibinfo {author} {\bibfnamefont
  {D.}~\bibnamefont {Flamm}}, \ and\ \bibinfo {author} {\bibfnamefont
  {F.}~\bibnamefont {Zimmermann}},\ }\bibfield  {title} {\enquote {\bibinfo
  {title} {Selective etching of ultrafast laser modified sapphire},}\ }in\
  \href@noop {} {\emph {\bibinfo {booktitle} {Laser Applications in
  Microelectronic and Optoelectronic Manufacturing (LAMOM) XXIV}}},\ Vol.\
  \bibinfo {volume} {10905}\ (\bibinfo {organization} {International Society
  for Optics and Photonics},\ \bibinfo {year} {2019})\ p.\ \bibinfo {pages}
  {109050F}\BibitemShut {NoStop}%
\bibitem [{\citenamefont {Rave}\ \emph {et~al.}(2021)\citenamefont {Rave},
  \citenamefont {Heiming}, \citenamefont {Szumny}, \citenamefont {Kaiser},
  \citenamefont {Kleiner},\ and\ \citenamefont {Flamm}}]{rave2021glass}%
  \BibitemOpen
  \bibfield  {author} {\bibinfo {author} {\bibfnamefont {H.}~\bibnamefont
  {Rave}}, \bibinfo {author} {\bibfnamefont {H.}~\bibnamefont {Heiming}},
  \bibinfo {author} {\bibfnamefont {P.}~\bibnamefont {Szumny}}, \bibinfo
  {author} {\bibfnamefont {M.}~\bibnamefont {Kaiser}}, \bibinfo {author}
  {\bibfnamefont {J.}~\bibnamefont {Kleiner}}, \ and\ \bibinfo {author}
  {\bibfnamefont {D.}~\bibnamefont {Flamm}},\ }\bibfield  {title} {\enquote
  {\bibinfo {title} {Glass tube cutting with aberration-corrected
  non-diffracting ultrashort laser pulses},}\ }\href@noop {} {\bibfield
  {journal} {\bibinfo  {journal} {Optical Engineering}\ }\textbf {\bibinfo
  {volume} {60}},\ \bibinfo {pages} {065105} (\bibinfo {year}
  {2021})}\BibitemShut {NoStop}%
\bibitem [{\citenamefont {Nisar}, \citenamefont {Li},\ and\ \citenamefont
  {Sheikh}(2013)}]{nisar2013laser}%
  \BibitemOpen
  \bibfield  {author} {\bibinfo {author} {\bibfnamefont {S.}~\bibnamefont
  {Nisar}}, \bibinfo {author} {\bibfnamefont {L.}~\bibnamefont {Li}}, \ and\
  \bibinfo {author} {\bibfnamefont {M.}~\bibnamefont {Sheikh}},\ }\bibfield
  {title} {\enquote {\bibinfo {title} {Laser glass cutting techniques—a
  review},}\ }\href@noop {} {\bibfield  {journal} {\bibinfo  {journal} {Journal
  of Laser Applications}\ }\textbf {\bibinfo {volume} {25}},\ \bibinfo {pages}
  {042010} (\bibinfo {year} {2013})}\BibitemShut {NoStop}%
\bibitem [{\citenamefont {Rubinsztein-Dunlop}\ \emph
  {et~al.}(2016)\citenamefont {Rubinsztein-Dunlop}, \citenamefont {Forbes},
  \citenamefont {Berry}, \citenamefont {Dennis}, \citenamefont {Andrews},
  \citenamefont {Mansuripur}, \citenamefont {Denz}, \citenamefont {Alpmann},
  \citenamefont {Banzer}, \citenamefont {Bauer} \emph
  {et~al.}}]{rubinsztein2016roadmap}%
  \BibitemOpen
  \bibfield  {author} {\bibinfo {author} {\bibfnamefont {H.}~\bibnamefont
  {Rubinsztein-Dunlop}}, \bibinfo {author} {\bibfnamefont {A.}~\bibnamefont
  {Forbes}}, \bibinfo {author} {\bibfnamefont {M.~V.}\ \bibnamefont {Berry}},
  \bibinfo {author} {\bibfnamefont {M.~R.}\ \bibnamefont {Dennis}}, \bibinfo
  {author} {\bibfnamefont {D.~L.}\ \bibnamefont {Andrews}}, \bibinfo {author}
  {\bibfnamefont {M.}~\bibnamefont {Mansuripur}}, \bibinfo {author}
  {\bibfnamefont {C.}~\bibnamefont {Denz}}, \bibinfo {author} {\bibfnamefont
  {C.}~\bibnamefont {Alpmann}}, \bibinfo {author} {\bibfnamefont
  {P.}~\bibnamefont {Banzer}}, \bibinfo {author} {\bibfnamefont
  {T.}~\bibnamefont {Bauer}},  \emph {et~al.},\ }\bibfield  {title} {\enquote
  {\bibinfo {title} {Roadmap on structured light},}\ }\href@noop {} {\bibfield
  {journal} {\bibinfo  {journal} {Journal of Optics}\ }\textbf {\bibinfo
  {volume} {19}},\ \bibinfo {pages} {013001} (\bibinfo {year}
  {2016})}\BibitemShut {NoStop}%
\bibitem [{\citenamefont {Woerdemann}(2012)}]{woerdemann2012structured}%
  \BibitemOpen
  \bibfield  {author} {\bibinfo {author} {\bibfnamefont {M.}~\bibnamefont
  {Woerdemann}},\ }\href@noop {} {\emph {\bibinfo {title} {Structured Light
  Fields: Applications in Optical Trapping, Manipulation, and Organisation}}}\
  (\bibinfo  {publisher} {Springer Science \& Business Media},\ \bibinfo {year}
  {2012})\BibitemShut {NoStop}%
\bibitem [{\citenamefont {Siviloglou}\ and\ \citenamefont
  {Christodoulides}(2007)}]{siviloglou2007accelerating}%
  \BibitemOpen
  \bibfield  {author} {\bibinfo {author} {\bibfnamefont {G.~A.}\ \bibnamefont
  {Siviloglou}}\ and\ \bibinfo {author} {\bibfnamefont {D.~N.}\ \bibnamefont
  {Christodoulides}},\ }\bibfield  {title} {\enquote {\bibinfo {title}
  {Accelerating finite energy {Airy} beams},}\ }\href@noop {} {\bibfield
  {journal} {\bibinfo  {journal} {Optics Letters}\ }\textbf {\bibinfo {volume}
  {32}},\ \bibinfo {pages} {979--981} (\bibinfo {year} {2007})}\BibitemShut
  {NoStop}%
\bibitem [{\citenamefont {Baumgartl}, \citenamefont {Mazilu},\ and\
  \citenamefont {Dholakia}(2008)}]{baumgartl2008optically}%
  \BibitemOpen
  \bibfield  {author} {\bibinfo {author} {\bibfnamefont {J.}~\bibnamefont
  {Baumgartl}}, \bibinfo {author} {\bibfnamefont {M.}~\bibnamefont {Mazilu}}, \
  and\ \bibinfo {author} {\bibfnamefont {K.}~\bibnamefont {Dholakia}},\
  }\bibfield  {title} {\enquote {\bibinfo {title} {Optically mediated particle
  clearing using {Airy} wavepackets},}\ }\href@noop {} {\bibfield  {journal}
  {\bibinfo  {journal} {Nature photonics}\ }\textbf {\bibinfo {volume} {2}},\
  \bibinfo {pages} {675--678} (\bibinfo {year} {2008})}\BibitemShut {NoStop}%
\bibitem [{\citenamefont {Mathis}\ \emph {et~al.}(2012)\citenamefont {Mathis},
  \citenamefont {Courvoisier}, \citenamefont {Froehly}, \citenamefont
  {Furfaro}, \citenamefont {Jacquot}, \citenamefont {Lacourt},\ and\
  \citenamefont {Dudley}}]{mathis2012micromachining}%
  \BibitemOpen
  \bibfield  {author} {\bibinfo {author} {\bibfnamefont {A.}~\bibnamefont
  {Mathis}}, \bibinfo {author} {\bibfnamefont {F.}~\bibnamefont {Courvoisier}},
  \bibinfo {author} {\bibfnamefont {L.}~\bibnamefont {Froehly}}, \bibinfo
  {author} {\bibfnamefont {L.}~\bibnamefont {Furfaro}}, \bibinfo {author}
  {\bibfnamefont {M.}~\bibnamefont {Jacquot}}, \bibinfo {author} {\bibfnamefont
  {P.-A.}\ \bibnamefont {Lacourt}}, \ and\ \bibinfo {author} {\bibfnamefont
  {J.~M.}\ \bibnamefont {Dudley}},\ }\bibfield  {title} {\enquote {\bibinfo
  {title} {Micromachining along a curve: Femtosecond laser micromachining of
  curved profiles in diamond and silicon using accelerating beams},}\
  }\href@noop {} {\bibfield  {journal} {\bibinfo  {journal} {Applied Physics
  Letters}\ }\textbf {\bibinfo {volume} {101}},\ \bibinfo {pages} {071110}
  (\bibinfo {year} {2012})}\BibitemShut {NoStop}%
\bibitem [{\citenamefont {Sohr}, \citenamefont {Thomas},\ and\ \citenamefont
  {Skupin}(2021)}]{sohr2021shaping}%
  \BibitemOpen
  \bibfield  {author} {\bibinfo {author} {\bibfnamefont {D.}~\bibnamefont
  {Sohr}}, \bibinfo {author} {\bibfnamefont {J.~U.}\ \bibnamefont {Thomas}}, \
  and\ \bibinfo {author} {\bibfnamefont {S.}~\bibnamefont {Skupin}},\
  }\bibfield  {title} {\enquote {\bibinfo {title} {Shaping convex edges in
  borosilicate glass by single pass perforation with an {Airy} beam},}\
  }\href@noop {} {\bibfield  {journal} {\bibinfo  {journal} {Optics Letters}\
  }\textbf {\bibinfo {volume} {46}},\ \bibinfo {pages} {2529--2532} (\bibinfo
  {year} {2021})}\BibitemShut {NoStop}%
\bibitem [{\citenamefont {Ungaro}\ and\ \citenamefont
  {Liu}(2021)}]{ungaro2021single}%
  \BibitemOpen
  \bibfield  {author} {\bibinfo {author} {\bibfnamefont {C.}~\bibnamefont
  {Ungaro}}\ and\ \bibinfo {author} {\bibfnamefont {A.}~\bibnamefont {Liu}},\
  }\bibfield  {title} {\enquote {\bibinfo {title} {Single-pass cutting of glass
  with a curved edge using ultrafast curving {Bessel} beams and oblong {Airy}
  beams},}\ }\href@noop {} {\bibfield  {journal} {\bibinfo  {journal} {Optics
  \& Laser Technology}\ }\textbf {\bibinfo {volume} {144}},\ \bibinfo {pages}
  {107398} (\bibinfo {year} {2021})}\BibitemShut {NoStop}%
\bibitem [{\citenamefont {Chremmos}\ \emph {et~al.}(2012)\citenamefont
  {Chremmos}, \citenamefont {Chen}, \citenamefont {Christodoulides},\ and\
  \citenamefont {Efremidis}}]{chremmos2012bessel}%
  \BibitemOpen
  \bibfield  {author} {\bibinfo {author} {\bibfnamefont {I.~D.}\ \bibnamefont
  {Chremmos}}, \bibinfo {author} {\bibfnamefont {Z.}~\bibnamefont {Chen}},
  \bibinfo {author} {\bibfnamefont {D.~N.}\ \bibnamefont {Christodoulides}}, \
  and\ \bibinfo {author} {\bibfnamefont {N.~K.}\ \bibnamefont {Efremidis}},\
  }\bibfield  {title} {\enquote {\bibinfo {title} {Bessel-like optical beams
  with arbitrary trajectories},}\ }\href@noop {} {\bibfield  {journal}
  {\bibinfo  {journal} {Optics Letters}\ }\textbf {\bibinfo {volume} {37}},\
  \bibinfo {pages} {5003--5005} (\bibinfo {year} {2012})}\BibitemShut {NoStop}%
\bibitem [{\citenamefont {Green}(2011)}]{green2011colour}%
  \BibitemOpen
  \bibfield  {author} {\bibinfo {author} {\bibfnamefont {D.~A.}\ \bibnamefont
  {Green}},\ }\bibfield  {title} {\enquote {\bibinfo {title} {A colour scheme
  for the display of astronomical intensity images},}\ }\href@noop {}
  {\bibfield  {journal} {\bibinfo  {journal} {arXiv preprint arXiv:1108.5083}\
  } (\bibinfo {year} {2011})}\BibitemShut {NoStop}%
\bibitem [{\citenamefont {Kumkar}\ \emph {et~al.}(2017)\citenamefont {Kumkar},
  \citenamefont {Kaiser}, \citenamefont {Kleiner}, \citenamefont {Flamm},
  \citenamefont {Grossmann}, \citenamefont {Bergner}, \citenamefont
  {Zimmermann},\ and\ \citenamefont {Nolte}}]{kumkar2017throughput}%
  \BibitemOpen
  \bibfield  {author} {\bibinfo {author} {\bibfnamefont {M.}~\bibnamefont
  {Kumkar}}, \bibinfo {author} {\bibfnamefont {M.}~\bibnamefont {Kaiser}},
  \bibinfo {author} {\bibfnamefont {J.}~\bibnamefont {Kleiner}}, \bibinfo
  {author} {\bibfnamefont {D.}~\bibnamefont {Flamm}}, \bibinfo {author}
  {\bibfnamefont {D.}~\bibnamefont {Grossmann}}, \bibinfo {author}
  {\bibfnamefont {K.}~\bibnamefont {Bergner}}, \bibinfo {author} {\bibfnamefont
  {F.}~\bibnamefont {Zimmermann}}, \ and\ \bibinfo {author} {\bibfnamefont
  {S.}~\bibnamefont {Nolte}},\ }\bibfield  {title} {\enquote {\bibinfo {title}
  {Throughput scaling by spatial beam shaping and dynamic focusing},}\ }in\
  \href@noop {} {\emph {\bibinfo {booktitle} {Laser Applications in
  Microelectronic and Optoelectronic Manufacturing (LAMOM) XXII}}},\ Vol.\
  \bibinfo {volume} {10091}\ (\bibinfo {organization} {International Society
  for Optics and Photonics},\ \bibinfo {year} {2017})\ p.\ \bibinfo {pages}
  {100910G}\BibitemShut {NoStop}%
\bibitem [{\citenamefont {Wyrowski}\ \emph {et~al.}(1994)\citenamefont
  {Wyrowski}, \citenamefont {van Esdonk}, \citenamefont {Zuidema},
  \citenamefont {Wadmann},\ and\ \citenamefont {Notenboom}}]{wyrowski1994use}%
  \BibitemOpen
  \bibfield  {author} {\bibinfo {author} {\bibfnamefont {F.}~\bibnamefont
  {Wyrowski}}, \bibinfo {author} {\bibfnamefont {H.}~\bibnamefont {van
  Esdonk}}, \bibinfo {author} {\bibfnamefont {R.~J.}\ \bibnamefont {Zuidema}},
  \bibinfo {author} {\bibfnamefont {S.}~\bibnamefont {Wadmann}}, \ and\
  \bibinfo {author} {\bibfnamefont {G.~J.}\ \bibnamefont {Notenboom}},\
  }\bibfield  {title} {\enquote {\bibinfo {title} {Use of diffractive optics in
  material processing},}\ }in\ \href@noop {} {\emph {\bibinfo {booktitle}
  {Diffractive and Holographic Optics Technology}}},\ Vol.\ \bibinfo {volume}
  {2152}\ (\bibinfo {organization} {International Society for Optics and
  Photonics},\ \bibinfo {year} {1994})\ pp.\ \bibinfo {pages}
  {139--144}\BibitemShut {NoStop}%
\bibitem [{\citenamefont {Flamm}\ \emph {et~al.}(2019)\citenamefont {Flamm},
  \citenamefont {Grossmann}, \citenamefont {Jenne}, \citenamefont {Zimmermann},
  \citenamefont {Kleiner}, \citenamefont {Kaiser}, \citenamefont {Hellstern},
  \citenamefont {Tillkorn},\ and\ \citenamefont {Kumkar}}]{flamm2019beam}%
  \BibitemOpen
  \bibfield  {author} {\bibinfo {author} {\bibfnamefont {D.}~\bibnamefont
  {Flamm}}, \bibinfo {author} {\bibfnamefont {D.~G.}\ \bibnamefont
  {Grossmann}}, \bibinfo {author} {\bibfnamefont {M.}~\bibnamefont {Jenne}},
  \bibinfo {author} {\bibfnamefont {F.}~\bibnamefont {Zimmermann}}, \bibinfo
  {author} {\bibfnamefont {J.}~\bibnamefont {Kleiner}}, \bibinfo {author}
  {\bibfnamefont {M.}~\bibnamefont {Kaiser}}, \bibinfo {author} {\bibfnamefont
  {J.}~\bibnamefont {Hellstern}}, \bibinfo {author} {\bibfnamefont
  {C.}~\bibnamefont {Tillkorn}}, \ and\ \bibinfo {author} {\bibfnamefont
  {M.}~\bibnamefont {Kumkar}},\ }\bibfield  {title} {\enquote {\bibinfo {title}
  {Beam shaping for ultrafast materials processing},}\ }in\ \href@noop {}
  {\emph {\bibinfo {booktitle} {Laser Resonators, Microresonators, and Beam
  Control XXI}}},\ Vol.\ \bibinfo {volume} {10904}\ (\bibinfo {organization}
  {International Society for Optics and Photonics},\ \bibinfo {year} {2019})\
  p.\ \bibinfo {pages} {109041G}\BibitemShut {NoStop}%
\bibitem [{\citenamefont {Zhu}\ \emph {et~al.}(2014)\citenamefont {Zhu},
  \citenamefont {Sun}, \citenamefont {Zhu}, \citenamefont {Chen}, \citenamefont
  {Gao}, \citenamefont {Ma},\ and\ \citenamefont {Zhang}}]{zhu2014three}%
  \BibitemOpen
  \bibfield  {author} {\bibinfo {author} {\bibfnamefont {L.}~\bibnamefont
  {Zhu}}, \bibinfo {author} {\bibfnamefont {M.}~\bibnamefont {Sun}}, \bibinfo
  {author} {\bibfnamefont {M.}~\bibnamefont {Zhu}}, \bibinfo {author}
  {\bibfnamefont {J.}~\bibnamefont {Chen}}, \bibinfo {author} {\bibfnamefont
  {X.}~\bibnamefont {Gao}}, \bibinfo {author} {\bibfnamefont {W.}~\bibnamefont
  {Ma}}, \ and\ \bibinfo {author} {\bibfnamefont {D.}~\bibnamefont {Zhang}},\
  }\bibfield  {title} {\enquote {\bibinfo {title} {Three-dimensional
  shape-controllable focal spot array created by focusing vortex beams
  modulated by multi-value pure-phase grating},}\ }\href@noop {} {\bibfield
  {journal} {\bibinfo  {journal} {Optics Express}\ }\textbf {\bibinfo {volume}
  {22}},\ \bibinfo {pages} {21354--21367} (\bibinfo {year} {2014})}\BibitemShut
  {NoStop}%
\bibitem [{\citenamefont {Gu}, \citenamefont {Li},\ and\ \citenamefont
  {Cao}(2014)}]{gu2014optical}%
  \BibitemOpen
  \bibfield  {author} {\bibinfo {author} {\bibfnamefont {M.}~\bibnamefont
  {Gu}}, \bibinfo {author} {\bibfnamefont {X.}~\bibnamefont {Li}}, \ and\
  \bibinfo {author} {\bibfnamefont {Y.}~\bibnamefont {Cao}},\ }\bibfield
  {title} {\enquote {\bibinfo {title} {Optical storage arrays: a perspective
  for future big data storage},}\ }\href@noop {} {\bibfield  {journal}
  {\bibinfo  {journal} {Light: Science \& Applications}\ }\textbf {\bibinfo
  {volume} {3}},\ \bibinfo {pages} {e177--e177} (\bibinfo {year}
  {2014})}\BibitemShut {NoStop}%
\bibitem [{\citenamefont {Jesacher}\ and\ \citenamefont
  {Booth}(2010)}]{jesacher2010parallel}%
  \BibitemOpen
  \bibfield  {author} {\bibinfo {author} {\bibfnamefont {A.}~\bibnamefont
  {Jesacher}}\ and\ \bibinfo {author} {\bibfnamefont {M.~J.}\ \bibnamefont
  {Booth}},\ }\bibfield  {title} {\enquote {\bibinfo {title} {Parallel direct
  laser writing in three dimensions with spatially dependent aberration
  correction},}\ }\href@noop {} {\bibfield  {journal} {\bibinfo  {journal}
  {Optics Express}\ }\textbf {\bibinfo {volume} {18}},\ \bibinfo {pages}
  {21090--21099} (\bibinfo {year} {2010})}\BibitemShut {NoStop}%
\bibitem [{\citenamefont {Valle}\ and\ \citenamefont
  {Cagigal}(2012)}]{valle2012analytic}%
  \BibitemOpen
  \bibfield  {author} {\bibinfo {author} {\bibfnamefont {P.~J.}\ \bibnamefont
  {Valle}}\ and\ \bibinfo {author} {\bibfnamefont {M.~P.}\ \bibnamefont
  {Cagigal}},\ }\bibfield  {title} {\enquote {\bibinfo {title} {Analytic design
  of multiple-axis, multifocal diffractive lenses},}\ }\href@noop {} {\bibfield
   {journal} {\bibinfo  {journal} {Optics Letters}\ }\textbf {\bibinfo {volume}
  {37}},\ \bibinfo {pages} {1121--1123} (\bibinfo {year} {2012})}\BibitemShut
  {NoStop}%
\bibitem [{\citenamefont {Grossmann}\ \emph {et~al.}(2016)\citenamefont
  {Grossmann}, \citenamefont {Reininghaus}, \citenamefont {Kalupka},
  \citenamefont {Kumkar},\ and\ \citenamefont
  {Poprawe}}]{grossmann2016transverse}%
  \BibitemOpen
  \bibfield  {author} {\bibinfo {author} {\bibfnamefont {D.}~\bibnamefont
  {Grossmann}}, \bibinfo {author} {\bibfnamefont {M.}~\bibnamefont
  {Reininghaus}}, \bibinfo {author} {\bibfnamefont {C.}~\bibnamefont
  {Kalupka}}, \bibinfo {author} {\bibfnamefont {M.}~\bibnamefont {Kumkar}}, \
  and\ \bibinfo {author} {\bibfnamefont {R.}~\bibnamefont {Poprawe}},\
  }\bibfield  {title} {\enquote {\bibinfo {title} {Transverse pump-probe
  microscopy of moving breakdown, filamentation and self-organized absorption
  in alkali aluminosilicate glass using ultrashort pulse laser},}\ }\href@noop
  {} {\bibfield  {journal} {\bibinfo  {journal} {Optics Express}\ }\textbf
  {\bibinfo {volume} {24}},\ \bibinfo {pages} {23221--23231} (\bibinfo {year}
  {2016})}\BibitemShut {NoStop}%
\bibitem [{\citenamefont {Bergner}\ \emph {et~al.}(2018)\citenamefont
  {Bergner}, \citenamefont {Seyfarth}, \citenamefont {Lammers}, \citenamefont
  {Ullsperger}, \citenamefont {D{\"o}ring}, \citenamefont {Heinrich},
  \citenamefont {Kumkar}, \citenamefont {Flamm}, \citenamefont
  {T{\"u}nnermann},\ and\ \citenamefont {Nolte}}]{bergner2018spatio}%
  \BibitemOpen
  \bibfield  {author} {\bibinfo {author} {\bibfnamefont {K.}~\bibnamefont
  {Bergner}}, \bibinfo {author} {\bibfnamefont {B.}~\bibnamefont {Seyfarth}},
  \bibinfo {author} {\bibfnamefont {K.}~\bibnamefont {Lammers}}, \bibinfo
  {author} {\bibfnamefont {T.}~\bibnamefont {Ullsperger}}, \bibinfo {author}
  {\bibfnamefont {S.}~\bibnamefont {D{\"o}ring}}, \bibinfo {author}
  {\bibfnamefont {M.}~\bibnamefont {Heinrich}}, \bibinfo {author}
  {\bibfnamefont {M.}~\bibnamefont {Kumkar}}, \bibinfo {author} {\bibfnamefont
  {D.}~\bibnamefont {Flamm}}, \bibinfo {author} {\bibfnamefont
  {A.}~\bibnamefont {T{\"u}nnermann}}, \ and\ \bibinfo {author} {\bibfnamefont
  {S.}~\bibnamefont {Nolte}},\ }\bibfield  {title} {\enquote {\bibinfo {title}
  {Spatio-temporal analysis of glass volume processing using ultrashort laser
  pulses},}\ }\href@noop {} {\bibfield  {journal} {\bibinfo  {journal} {Applied
  Optics}\ }\textbf {\bibinfo {volume} {57}},\ \bibinfo {pages} {4618--4632}
  (\bibinfo {year} {2018})}\BibitemShut {NoStop}%
\bibitem [{\citenamefont {Jansen}, \citenamefont {Budnicki},\ and\
  \citenamefont {Sutter}(2018)}]{jansen2018pulsed}%
  \BibitemOpen
  \bibfield  {author} {\bibinfo {author} {\bibfnamefont {F.}~\bibnamefont
  {Jansen}}, \bibinfo {author} {\bibfnamefont {A.}~\bibnamefont {Budnicki}}, \
  and\ \bibinfo {author} {\bibfnamefont {D.}~\bibnamefont {Sutter}},\
  }\bibfield  {title} {\enquote {\bibinfo {title} {Pulsed lasers for industrial
  applications: Fiber, slab and thin-disk: Ultrafast laser technology for every
  application},}\ }\href@noop {} {\bibfield  {journal} {\bibinfo  {journal}
  {Laser Technik Journal}\ }\textbf {\bibinfo {volume} {15}},\ \bibinfo {pages}
  {46--49} (\bibinfo {year} {2018})}\BibitemShut {NoStop}%
\bibitem [{\citenamefont {Herman}, \citenamefont {Marjoribanks},\ and\
  \citenamefont {Oettl}(2003)}]{herman2003burst}%
  \BibitemOpen
  \bibfield  {author} {\bibinfo {author} {\bibfnamefont {P.~R.}\ \bibnamefont
  {Herman}}, \bibinfo {author} {\bibfnamefont {R.}~\bibnamefont
  {Marjoribanks}}, \ and\ \bibinfo {author} {\bibfnamefont {A.}~\bibnamefont
  {Oettl}},\ }\href@noop {} {\enquote {\bibinfo {title} {Burst-ultrafast laser
  machining method},}\ } (\bibinfo {year} {Apr.~22 2003}),\ \bibinfo {note}
  {{US Patent} 6,552,301}\BibitemShut {NoStop}%
\bibitem [{\citenamefont {Kerse}\ \emph {et~al.}(2016)\citenamefont {Kerse},
  \citenamefont {Kalayc{\i}o{\u{g}}lu}, \citenamefont {Elahi}, \citenamefont
  {{\c{C}}etin}, \citenamefont {Kesim}, \citenamefont {Ak{\c{c}}aalan},
  \citenamefont {Yava{\c{s}}}, \citenamefont {A{\c{s}}{\i}k}, \citenamefont
  {{\"O}ktem}, \citenamefont {Hoogland} \emph {et~al.}}]{kerse2016ablation}%
  \BibitemOpen
  \bibfield  {author} {\bibinfo {author} {\bibfnamefont {C.}~\bibnamefont
  {Kerse}}, \bibinfo {author} {\bibfnamefont {H.}~\bibnamefont
  {Kalayc{\i}o{\u{g}}lu}}, \bibinfo {author} {\bibfnamefont {P.}~\bibnamefont
  {Elahi}}, \bibinfo {author} {\bibfnamefont {B.}~\bibnamefont {{\c{C}}etin}},
  \bibinfo {author} {\bibfnamefont {D.~K.}\ \bibnamefont {Kesim}}, \bibinfo
  {author} {\bibfnamefont {{\"O}.}~\bibnamefont {Ak{\c{c}}aalan}}, \bibinfo
  {author} {\bibfnamefont {S.}~\bibnamefont {Yava{\c{s}}}}, \bibinfo {author}
  {\bibfnamefont {M.~D.}\ \bibnamefont {A{\c{s}}{\i}k}}, \bibinfo {author}
  {\bibfnamefont {B.}~\bibnamefont {{\"O}ktem}}, \bibinfo {author}
  {\bibfnamefont {H.}~\bibnamefont {Hoogland}},  \emph {et~al.},\ }\bibfield
  {title} {\enquote {\bibinfo {title} {Ablation-cooled material removal with
  ultrafast bursts of pulses},}\ }\href@noop {} {\bibfield  {journal} {\bibinfo
   {journal} {Nature}\ }\textbf {\bibinfo {volume} {537}},\ \bibinfo {pages}
  {84--88} (\bibinfo {year} {2016})}\BibitemShut {NoStop}%
\bibitem [{\citenamefont {Jenne}\ \emph
  {et~al.}(2018{\natexlab{b}})\citenamefont {Jenne}, \citenamefont
  {Zimmermann}, \citenamefont {Flamm}, \citenamefont {Gro{\ss}mann},
  \citenamefont {Kleiner}, \citenamefont {Kumkar},\ and\ \citenamefont
  {Nolte}}]{jenne2018multi}%
  \BibitemOpen
  \bibfield  {author} {\bibinfo {author} {\bibfnamefont {M.}~\bibnamefont
  {Jenne}}, \bibinfo {author} {\bibfnamefont {F.}~\bibnamefont {Zimmermann}},
  \bibinfo {author} {\bibfnamefont {D.}~\bibnamefont {Flamm}}, \bibinfo
  {author} {\bibfnamefont {D.}~\bibnamefont {Gro{\ss}mann}}, \bibinfo {author}
  {\bibfnamefont {J.}~\bibnamefont {Kleiner}}, \bibinfo {author} {\bibfnamefont
  {M.}~\bibnamefont {Kumkar}}, \ and\ \bibinfo {author} {\bibfnamefont
  {S.}~\bibnamefont {Nolte}},\ }\bibfield  {title} {\enquote {\bibinfo {title}
  {Multi pulse pump-probe diagnostics for development of advanced transparent
  materials processing},}\ }\href@noop {} {\bibfield  {journal} {\bibinfo
  {journal} {Journal of Laser Micro Nanoengineering}\ }\textbf {\bibinfo
  {volume} {13}},\ \bibinfo {pages} {273--279} (\bibinfo {year}
  {2018}{\natexlab{b}})}\BibitemShut {NoStop}%
\bibitem [{\citenamefont {Gottmann}\ \emph {et~al.}(2017)\citenamefont
  {Gottmann}, \citenamefont {Hermans}, \citenamefont {Repiev},\ and\
  \citenamefont {Ortmann}}]{gottmann2017selective}%
  \BibitemOpen
  \bibfield  {author} {\bibinfo {author} {\bibfnamefont {J.}~\bibnamefont
  {Gottmann}}, \bibinfo {author} {\bibfnamefont {M.}~\bibnamefont {Hermans}},
  \bibinfo {author} {\bibfnamefont {N.}~\bibnamefont {Repiev}}, \ and\ \bibinfo
  {author} {\bibfnamefont {J.}~\bibnamefont {Ortmann}},\ }\bibfield  {title}
  {\enquote {\bibinfo {title} {Selective laser-induced etching of {3D}
  precision quartz glass components for microfluidic applications—up-scaling
  of complexity and speed},}\ }\href@noop {} {\bibfield  {journal} {\bibinfo
  {journal} {Micromachines}\ }\textbf {\bibinfo {volume} {8}},\ \bibinfo
  {pages} {110} (\bibinfo {year} {2017})}\BibitemShut {NoStop}%
\bibitem [{\citenamefont {Hermans}, \citenamefont {Gottmann},\ and\
  \citenamefont {Riedel}(2014)}]{hermans2014selective}%
  \BibitemOpen
  \bibfield  {author} {\bibinfo {author} {\bibfnamefont {M.}~\bibnamefont
  {Hermans}}, \bibinfo {author} {\bibfnamefont {J.}~\bibnamefont {Gottmann}}, \
  and\ \bibinfo {author} {\bibfnamefont {F.}~\bibnamefont {Riedel}},\
  }\bibfield  {title} {\enquote {\bibinfo {title} {Selective, laser-induced
  etching of fused silica at high scan-speeds using {KOH}.}}\ }\href@noop {}
  {\bibfield  {journal} {\bibinfo  {journal} {Journal of Laser
  Micro/Nanoengineering}\ }\textbf {\bibinfo {volume} {9}} (\bibinfo {year}
  {2014})}\BibitemShut {NoStop}%
\end{thebibliography}%

\end{document}